\documentclass[showpacs,aps,pre,superscriptaddress,twocolumn]{revtex4}

\usepackage{graphicx}
\usepackage{amsfonts}
\usepackage{amssymb}
\usepackage{amsmath}
\usepackage{color,verbatim} 

\graphicspath{{./figures/}{./figures_rcg/}}

\begin{document}

\title{Superfluid vortex multipoles and soliton stripes on a torus}

\author{J. D'Ambroise}
\affiliation{ Department of Mathematics, Computer \& Information Science, State University of New York (SUNY) College at Old Westbury, Westbury, NY, 11568, USA}

\author{R. Carretero-Gonz{\'a}lez}
\affiliation{Nonlinear Dynamical Systems
Group,\footnote{\texttt{URL}: http://nlds.sdsu.edu}
Computational Sciences Research Center, and
Department of Mathematics and Statistics,
San Diego State University, San Diego, California 92182-7720, USA}

\author{P. Schmelcher}
\affiliation{Center for Optical Quantum Technologies, Department of Physics, 
University of Hamburg, Luruper Chaussee 149, 22761 Hamburg Germany}
\affiliation{The Hamburg Centre for Ultrafast Imaging,
University of Hamburg, Luruper Chaussee 149, 22761 Hamburg,
Germany}

\author{P.G. Kevrekidis}
\affiliation{Department of Mathematics and Statistics, University of Massachusetts,
Amherst, MA, 01003, USA}

\begin{abstract}
We study the existence, stability, and dynamics of vortex dipole and
quadrupole configurations in the nonlinear Schr\"odinger (NLS) equation
on the surface of a torus.
For this purpose we use, in addition to the full two-dimensional NLS
on the torus, a recently derived [Phys.~Rev.~A {\bf 101}, 053606
(2021)] reduced point-vortex particle model which is shown
to be in excellent agreement with the full NLS evolution.
Horizontal, vertical, and diagonal stationary vortex dipoles are identified and
continued along the torus aspect ratio and the chemical potential of the solution.
Windows of stability for these solutions are identified.
We also investigate stationary vortex quadrupole configurations. 
After eliminating similar solutions induced by invariances and symmetries, 
we find a total of 16 distinct configurations
ranging from horizontal and vertical aligned quadrupoles, to rectangular and 
rhomboidal quadrupoles, to trapezoidal and irregular quadrupoles.
The stability for the least unstable and, potentially, stable quadrapole solutions is 
monitored both at the NLS and the reduced model levels.
Two quadrupole configurations are found to be stable on small windows of the 
torus aspect ratio and a handful of quadrupoles are found to be very
weakly unstable for relatively large parameter windows.
Finally, we briefly study the dark soliton stripes
and their connection, through a series of bifurcation cascades, with steady
state vortex configurations.
\end{abstract}
\pacs{}

\maketitle

\section{Introduction}
\label{sec:intro}

The study of atomic Bose-Einstein condensates (BECs) offers a pristine
setting to explore the interplay of nonlinear dynamical
phenomena and quantum mechanical effects~\cite{stringari, pethick,
  siambook}.
A major thrust of associated experimental and theoretical
efforts has consisted of the exploration of coherent structures
supported by the interplay of effective nonlinearity and dispersion
in such systems, both at the mean-field level but also
beyond~\cite{tsatsos}. More specifically, relevant studies
and a wide range of experiments have focused on bright
solitons~\cite{experiment2,expb1,expb2,expb3} in attractively-interacting
condensates, dark solitons in self-repulsive
species~\cite{experiment,experiment1,becker,markus,engels,becker2,markus2,djf},
gap solitons~\cite{gap}, as well as multi-component
structures~\cite{revip}. While the above have been prototypically
one-dimensional states, higher dimensional structures such as
vortices~\cite{fetter1,fetter2}
and vortex rings~\cite{jeff,komineas_rev} have also attracted
significant attention in their own right.

Naturally, this activity has been mostly focused in the prototypical
settings of parabolic (but also often periodic) traps in one-
and higher dimensions, that have been the typical settings
of experiments so far~\cite{stringari, pethick,siambook}.
However, recent years, have seen a surge of
activity as concerns the exploration of BECs in 2D surfaces.
In the last year alone, multiple papers explored the dynamics
of vortices and vortex-antivortex papers in spherical surface,
shell-shaped systems~\cite{lahnert,fetter3}, following up on
earlier related work not only on spherical surfaces~\cite{tononi},
but also on cylindrical surfaces, planar annuli and sectors, as well
as cones~\cite{fetter:cylinder,fetter:cone}. The topic of the dynamics
of vortices on curved surfaces is one that bears considerable
history motivated by a variety of settings in fluid~\cite{pknewton}
and superfluid~\cite{vitelli} physics. In terrestrial BEC settings,
the realization of such experiments suffers from aspects such as
the gravitational sag. However, the recent activity in the
newly launched Cold Atomic Laboratory (CAL)
aboard the International Space Station seems to hold considerable
promise in this direction and indeed is specifically aiming to
implement a hollow bubble geometry~\cite{lenn26,lenn27}. This, in turn, paves the
way for the broader study of pattern dynamics (including of
topologically charged states such as vortices) in nontrivial
geometry and topology-featuring setups. It should be also noted
that this is  in addition to the remarkable recent developments
towards confining and manipulating atoms via adiabatic potentials,
which, in turn, can also lead to a diverse variety of traps for
ultracold atoms; see, e.g., Ref.~\cite{garraway} for a relevant review.

In the present work, our platform of choice will be the
surface a torus, i.e., the simplest compact and multiply connected surface.
This is motivated by the above developments, the torus' nontrivial topological
structure, by the proposal by experimental groups of the
realization of an optical lattice on its surface~\cite{porto} and by
the recent formulation of the effective vortex particle dynamics
on its surface~\cite{fetter:torus}. More concretely,
we extend our earlier considerations in
the realm of bright solitary waves~\cite{torus:bright} to the
self-repulsive condensate setting and the primarily vortical
(but also dark soliton stripe) set of structures that can arise therein,
while also respecting the periodicity of the torus in both of its angular
directions. The fundamental work of Ref.~\cite{fetter:torus} has
set the stage by providing a description at the level of
ordinary differential equations (ODEs) for the vortex particles. Yet,
the latter work still left many questions unanswered. For instance,
while this description is applicable at large chemical potentials,
it is useful to explore the nature of the existence, stability and
dynamics of multi-vortex structures as a function of the
chemical potential (which is also a proxy for the atom number),
but also as a function of the torus geometric parameters such
as the ratio of the minor to the major axis. Furthermore, while
the ODEs were derived, the potential equilibria of those and
the associated stability and phase portraits were not explored
even for the most prototypical case of a vortex pair. Indeed, there
are further significant multi-vortex configurations that
are relevant to consider such as the vortex quadrupoles. Additionally,
as we will see below, the vortex patterns also bear connections
(through their bifurcations) to states involving dark solitonic
stripes that are of interest in their own right.
Last but by no means least, it is
also particularly meaningful to compare the ODE results with direct
partial differential equation (PDE) simulations,
to explore the validity and also potential limitations of the approach. 

More concretely, our work is organized as follows. In Sec.~\ref{sec:thy} we introduce
the original, full, spatio-temporal NLS model on the torus and briefly
review (for completeness) the main aspects of its 
reduction to an effective point-vortex model, as obtained in Ref.~\cite{fetter:torus}.
Section~\ref{sec:num} presents the bulk of our results by studying the
existence of vortex- and stripe-bearing solutions, their stability and dynamics.
In particular, in Sec.~\ref{sec:num:dip} we exhaustively analyze the
existence, stability, and dynamics for vortex dipole configurations.
We find a total of 4 different stationary dipole solutions, two of which are fully stable.
We also extend these solutions by adding extra phase windings along the
toroidal and poloidal directions.
Section~\ref{sec:num:quad} is devoted to the study of quadrupole
configurations where we identify a total of 16 distinct ones. A couple
of these quadrupoles are found to be stable within small parameter windows
while a handful of quadrupoles are found to be very weakly unstable
for relatively large parameter windows.
In Sec.~\ref{sec:num:DSSs} we briefly study the existence and
stability of dark soliton configurations and connect some of them,
via bifurcation cascades, with steady state vortex patterns.
Finally, a summary and conclusions of our work, together with some
possible avenues for future research, are given in Sec.~\ref{sec:conclu}.

\section{Model and Theoretical Setup}
\label{sec:thy}

\subsection{Spatiotemporal model}
\label{sec:spatio}

\begin{figure}
\hspace{-0.0in}\includegraphics[width=0.7\columnwidth]{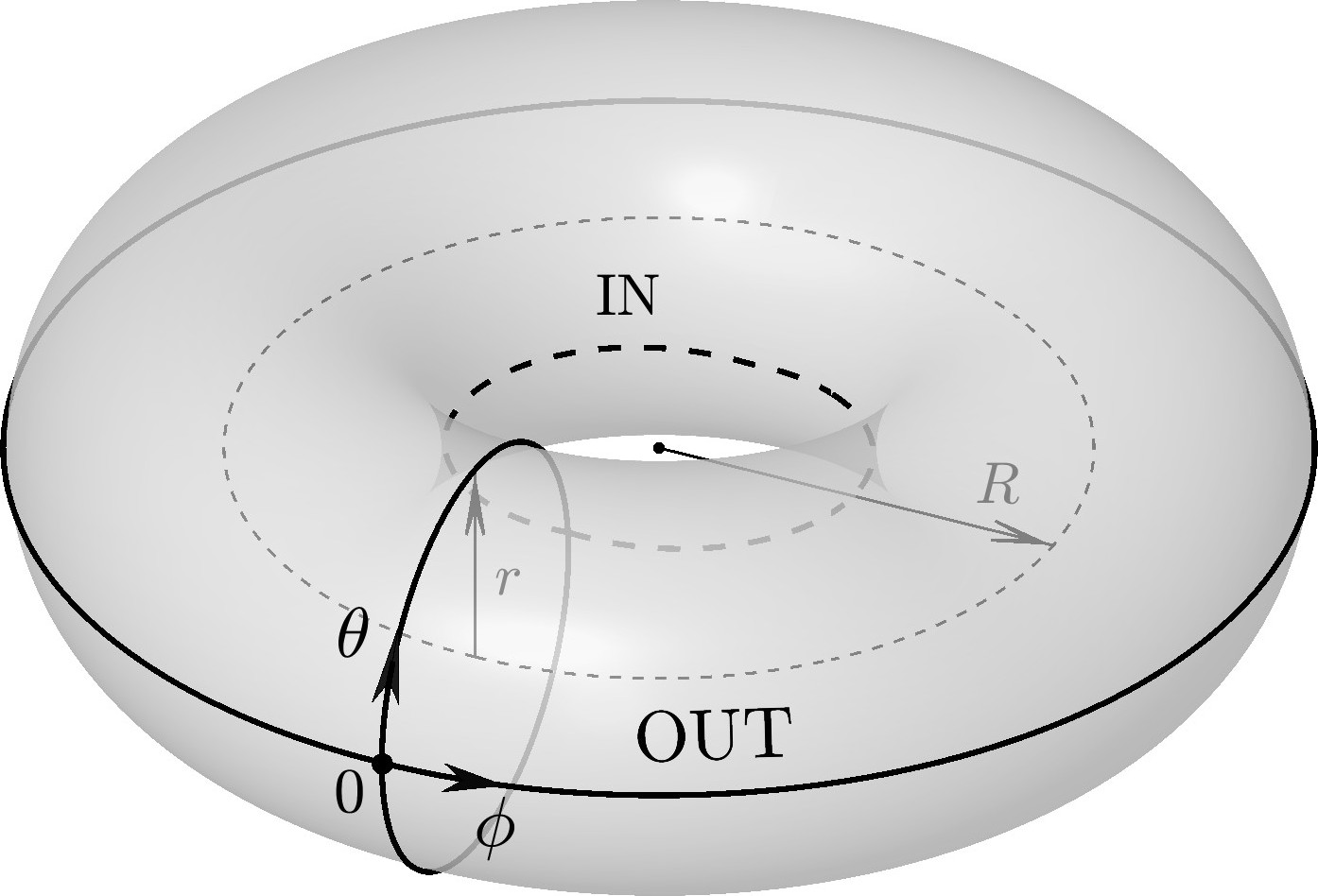} 
\caption{
Torus with major radius $R$ and a minor radius $r$
and the corresponding toroidal $\phi$ and poloidal $\theta$ angles.
}
\label{fig:torus3D}
\end{figure}

The NLS solutions that we study live on the torus centered at the origin.
The torus has a major (toroidal) radius $R$ and a minor (poloidal) radius $r$ 
such that $R>r$. The torus coordinates in 3D space are parametrized by
\begin{equation}
\label{eq:3Dtorus}
\left\{
\begin{array}{rcl}
X &=& [R+r\cos(\theta)]\cos(\phi), 
\\[1.0ex]
Y &=& [R+r\cos(\theta)]\sin(\phi), 
\\[1.0ex]
Z &=& r\sin(\theta).
\end{array}
\right.
\end{equation}
It is useful to define the angular coordinates on the torus as follows:
the toroidal angle is denoted by $\phi\in[0,2\pi]$ and poloidal 
angle $\theta\in[0,2\pi]$.  
In particular, we choose our toroidal axis such that $\theta=0$ corresponds
to the outermost ring of the torus while $\theta=\pi$ corresponds to the 
innermost ring; see Fig.~\ref{fig:torus3D}.
Hence, solutions along these rings will be dubbed below as
outer and inner respectively.
On the surface of this torus, in the absence of any external potentials,
the 2D NLS wavefunction $\psi=\psi(\phi,\theta,t)$ is described by the 
adimensionalized spatiotemporal model
\begin{eqnarray}
\label{eq:nls}
i\psi_t  = -\frac{1}{2}\Delta\psi - \sigma |\psi|^2\psi ,
\end{eqnarray}
where $\sigma =+1$ and $\sigma=-1$
correspond, respectively, to the focusing (attractive) and 
defocusing (repulsive) cases. 
In the present study we exclusively examine the defocusing case that admits
`dark' structures (i.e., structures on top of a finite background)
such as dark soliton stripes and vortices, hence in our computations
hereafter $\sigma=-1$.
Central to the description of the NLS on a torus is, of course, the corresponding
Laplace-Beltrami operator that takes the form (see, e.g., Ref.~\cite{glowinski})
\begin{eqnarray}
  \Delta =  \frac{1}{r^2}\partial_{\theta\theta} - \frac{\sin\theta}{r(R+r\cos\theta)} \partial_\theta + \frac{1}{(R+r\cos\theta)^2}\partial_{\phi\phi}.
  \label{lap}
  \end{eqnarray}
In what follows, we define the torus aspect ratio $\alpha\equiv r/R$.

\subsection{Steady states and Stability}
\label{sec:steady_stab}

In Sec.~\ref{sec:num} we will construct and study several steady state solutions for
the above NLS on the torus. These steady states (standing waves)
are found by separating space and 
time variables according to $\psi(\phi,\theta,t) = \varphi_0(\phi,\theta)\, e^{-i\mu t}$,
where $\mu$ is often referred to as the chemical potential and corresponds
to the temporal frequency of the solution (as well as to the density of its background). 
Thus, the steady state NLS for $\varphi_0(\phi,\theta)$, parametrized by $\mu$, reads
\begin{eqnarray}
\label{stateq}
-\frac{1}{2}\Delta \varphi_0  - \sigma |\varphi_0|^2\varphi_0 -\mu \varphi_0 = 0.
\end{eqnarray}
After suitably identifying steady states (for numerical details, see below),
it is relevant to study their dynamical
stability properties by extracting the corresponding stability spectra.
Therefore, we follow perturbed solutions starting by a steady state solution 
$\varphi_0$ as per Eq.~(\ref{stateq}) and perturb it
with an infinitesimal perturbation according to
\begin{eqnarray}
  \psi &=& \left\{\varphi_0 + \varepsilon
\left[ae^{\lambda t} + b^* e^{\lambda^* t}\right]\right\}
e^{-i\mu t},
  \label{perturb}
\end{eqnarray}
where $a=a(\phi,\theta)$ and $b=b(\phi,\theta)$ correspond to the spatial eigenmodes
with eigenvalue $\lambda$.
Then, the so-called Bogolyubov-de Gennes (BdG) stability spectra is obtained
by solving the linearization equation (i.e., to order $\varepsilon^1$):
\begin{equation}
\label{eq:mat}
\left[ \begin{array}{cc} M_1 & M_2 \\ -M_2^* & -M_1^* \end{array} \right] \left[ \begin{array}{c} a\\ b \end{array} \right] =  -i\lambda \left[ \begin{array}{c} a\\ b \end{array} \right],
\end{equation}
where $M_1 = \frac{1}{2} \Delta + \mu + 2\sigma |\varphi_0|^2$ and $M_2 = \sigma \varphi_0^2$.   
By construction, the spectrum obtained from Eq.~(\ref{perturb}) will respect the
Hamiltonian symmetry such that if $\lambda$ is an eigenvalue, so is $-\lambda$,
$\lambda^*$, and $-\lambda^*$ where the asterisk stands for complex conjugation.
Therefore, any eigenvalue such that Re$(\lambda)> 0$ will correspond to
an instability ---an exponential instability if Im$(\lambda)=0$ and
an oscillatory instability if Im$(\lambda)\not= 0$. In the latter
case, the exponential growth is also accompanied by an oscillatory
dynamics of the solution.

In the present work we focus primarily on stationary solutions composed of multi-vortex 
configurations. Similar to the focusing case of Ref.~\cite{torus:bright}, there will 
be special vortex locations on the torus corresponding to stationary configurations.
It is important to mention that, due to the periodic nature of the domain, only
configurations with  zero total charge are allowed. Therefore, only configurations
with the same number of positively and negatively charge vortices are
possible within the toroidal geometry. 
Therefore, we will focus on the lowest order ones that correspond to vortex 
{\em dipoles} and {\em quadrupoles}. As indicated also in the
introduction, even for the former there are a lot of important features
to explore including their stationary configurations and associated
stability, while the latter have not been considered at all
previously, to the best of our knowledge.
The periodicity of the domain also allows for the existence of dark soliton stripe
configurations provided they appear in pairs to allow for the individual
$\pi$ phase jumps of each dark soliton to accumulate to a whole $2\pi$ phase jump.
Nonetheless, as we will see in Sec.~\ref{sec:num:DSSs}, configurations with
odd number of dark soliton stripes are also possible by adding extra phase
windings (perpendicular to the stripes) to respect the periodicity.
Finally, it is also relevant to note that the Laplacian operator in Eq.~(\ref{lap}) 
is translationally invariant along the toroidal $\phi$-direction. Thus,
steady state solutions will generate an entire family of possible $\phi$-translates
---unless the steady states is already homogeneous in the $\phi$-direction
such as all horizontal (toroidal) dark soliton stripe configurations.

\subsection{Reduced point-vortex particle model}
\label{sec:ptvortex}

In tandem with the procurement of steady states and their characterization (stability),
we will study the corresponding elements in the dynamically reduced model where the vortices 
are considered as point-particles, as per the fundamental work of Ref.~\cite{fetter:torus}.
Note that, in comparison to the latter, we are using here adimensional variables which
correspond to the models in these works with $\hbar=M=1$ where $M$ is the 
mass of the particles forming the atomic BEC.
The model is  cast as a set of ODEs on the vortex positions.
This point-vortex model assumes no internal (density) structure for the vortices and
that the only effects come from vortex-vortex phase interactions and, importantly,
the curvature effects from the toroidal substrate where they are embedded.
Such point-vortex models have been shown to be accurate in the (sufficiently) large 
$\mu$-limit~\cite{fetter1,siambook}.
Indeed, what we will conclude here, as well, is that when
$R$ and $r$ are of order unity, a value of $\mu=5$ (and sometimes even as low as $\mu=1$) 
is sufficiently large such that the
reduced model gives an accurate static (steady states), stability, and 
dynamical representation of the full NLS model on a torus.

In the large $\mu$-limit, one only considers the phase of the vortices
and how the superposition of the phases from all other vortices
advects
the position of each vortex through the identification of the local fluid
velocity as the gradient of the wavefunction's phase at that point.
The key toward setting up the point-vortex model on the torus lies in taking 
into consideration two crucial effects that are absent in the standard 
NLS model on a flat and infinite domain. Namely, (i) the effects of the 
periodic boundary conditions and (ii) the effects of the torus' curvature.
The periodic boundary conditions are accounted for by placing `ghost' (or mirror) 
vortices outside the domain accounting for the effects that a particular
vortex has on itself through the boundaries as well as the effects of the
other vortices through the boundaries. 
This cumulative process results in an infinite sum for these contributions
that can be represented in the form of
Weierstrass functions or in terms of Jacobi-$\theta$ 
functions; see Ref\.~\cite{stremler:JFM99} and references therein.
On the other hand, the effect of curvature from the torus can be more
conveniently captured by expressing the system in isothermal coordinates.
Following Ref.~\cite{fetter:torus}, one defines the isothermal 
coordinates $(u,v)$ related to the toroidal $(\phi,\theta)$ ones 
through~\cite{kirchhoff}
\begin{equation}
\label{eq:iso-tor}
\left\{
\begin{array}{rcl}
\phi&=&\displaystyle\frac{u}{c},
\\[1.0ex]
\theta&=&\displaystyle2\tan^{-1}\left[ \sqrt{\frac{R+r}{R-r}}\tan\left(\frac{v}{2r}\right)\right],
\end{array}
\right.
\end{equation}
where $c\equiv\sqrt{R^2-r^2}$.
Then, it is possible to show that these new $(u,v)$ coordinates, with squared 
line element $ds^2=\Lambda^2(du^2+dv^2)$, are indeed isothermal (i.e.,
local ones in which the metric is conformal to the Euclidean) if the
local scale factor satisfies
\begin{equation}
\label{eq:scalefact}
\Lambda=\frac{c}{R-r\cos(v/r)}.
\end{equation}
Note that the scale factor only depends on the poloidal location, namely $\Lambda=\Lambda(v)$, 
since the system is translationally invariant along the toroidal direction.
The isothermal coordinates $(u,v)$ are then defined on the periodic
rectangle $[-\pi c,\pi c]\times[-\pi r,\pi r]$.

Taking the periodic and curvature effects and defining the complex coordinate
$w=u+iv$, the work of Ref.~\cite{fetter:torus} gives the explicit form for the
wavefunction's phase $\Phi$ associated to a set of vortex dipoles composed 
of a total zero charge configuration of vortices with charges $q_n$ at 
(isothermal) positions $w_n$.  Namely, $\Phi(w)={\rm Im}({\cal F}(w))$ where
${\cal F}(w)=\sum_n q_n F(w,w_n)$ and
\begin{equation}
\label{eq:F}
F(w,w_n) \equiv \ln\left[\vartheta_1\left(\frac{w-w_n}{2c},p\right)\right]
-\frac{{\rm Re}(w_n)}{2\pi r c}w,
\end{equation}
and $\vartheta_1(w,p)$ is the first Jacobi-$\theta$ function evaluated at $w$
with nome $p\equiv e^{-\pi r/c}$ ($0<p<1$). The first Jacobi-$\theta$ function
may be written as the following infinite sum
\begin{equation}
\label{eq:Jacobi}
\vartheta_1(w,p)=2 p^{1/4}\sum_{n=0}^\infty{(-1)^n p^{n(n+1)}\sin[(2n+1)w].}
\end{equation}
Note that this implementation of $\vartheta_1$ requires, for typical numerical
values used in this manuscript, less than a dozen terms for this infinite sum
to converge to machine (double) precision.
From this overall phase imparted by all the vortex dipoles, one can explicitly
write equations of motion for the individual vortices through the fluid velocity
that, in turn, is equivalent to the gradient of the phase. 
Thus, the $n$-th vortex will experience a velocity $V_n$
given by the gradient of the phase imprinted by the {\em other} vortices.
Expressing the velocity in complex coordinates $V_n=\dot u_n+i\dot v_n$ yields 
$\dot u_n={\rm Im}(\Omega_n)/\Lambda(v_n)$ and 
$\dot v_n={\rm Re}(\Omega_n)/\Lambda(v_n)$ where
\begin{equation}
\label{eq:ODEvel}
\Omega_n = \frac{1}{\Lambda(v_n)}
\left[\sum_{m\not= n}
q_m f(w_n,w_m) 
+ i \frac{q_n}{2}\frac{\Lambda'(v_n)}{\Lambda(v_n)}
- \frac{q_n u_n}{2\pi r c}
\right],
\end{equation}
where
\begin{equation}
\label{eq:f}
f(w,w_n)\equiv \partial_w F(w,w_n).
\end{equation}
Note that the sum excludes self-interacting terms (i.e., $m\not= n$).
By construction, the function $f(w,w_n)$ is periodic in both the imaginary 
(vertical) and real (horizontal) directions. The vertical periodicity is 
captured from the explicit periodicity of the Jacobi-$\theta$ function itself, 
while the horizontal periodicity is achieved by judiciously adding a linear term 
in the horizontal direction, see last term in Eq.~(\ref{eq:F}), that ensures 
continuity of the velocity in the periodic domain~\cite{fetter:torus}.
In the next section, the point-vortex model, cast through the explicit velocity
formulation of Eq.~(\ref{eq:ODEvel}), is validated against numerical results from
the full NLS (\ref{eq:nls}). This point-vortex model will also be instrumental in
finding stationary dipole and quadrupole solutions.

\section{Numerical Results}
\label{sec:num}

In order to find branches of solutions as the system parameters are varied
it is usually sufficient to find a single element of the branch and then
apply numerical continuation to extend each branch over these parameters
and study their existence and stability as the system parameters (mostly the 
chemical potential $\mu$ and the torus aspect ratio $\alpha$) are varied.
Thus, let us now leverage the results from the previous section to find
particular stationary vortex configurations (dipoles and quadrupoles)
for the reduced ODE model (\ref{eq:ODEvel}) and from there construct
approximate steady states for the full NLS (\ref{eq:nls}) that can
be used with a fixed-point iteration scheme (cf.~Newton's method) to 
find numerically exact steady states.
Once a particular steady state of (even) $N$ vortices with charges $q_n$ and
locations $(u_n,v_n)$ is identified in the reduced ODE model,
we construct an approximate initial wavefunction seed $\varphi_0$ by 
`superimposing' individual vortex-like guesses as follows 
\begin{equation}
\label{eq:seed}
  \varphi_0(u,v)=\sqrt{\mu}\cdot e^{i\Phi(u,v)} \cdot \Pi_{n=1}^N A(u-u_n,v-v_n),
\end{equation}
containing the following ingredients.
(i) The background level is fixed so that,
away from vortices, the density $|\varphi_0|^2$ tends to $\mu$.
(ii) The global phase $\Phi(u,v)$ is prescribed by the ODE model as per
$\Phi(w)={\rm Im}(\sum_n q_n F(w,w_n))$ with $F$ defined in Eq.~(\ref{eq:F}).
(iii) Each vortex (absolute value) profile is approximated by 
\begin{equation}
\label{eq:amp} 
A(u,v)=\tanh(\sqrt{\mu}\sqrt{u^2+v^2}),
\end{equation}
centered at each of the vortex locations $(u_n,v_n)$.
In a similar vein, one can also construct approximate dark soliton
stripe solutions as follows
\begin{equation}
  \varphi_0(u,v)=\sqrt{\mu}\cdot  \Pi_{n=1}^N B(u_n,v_n)
\end{equation}
where $B(u_n,v_n)=A(u-u_n,0)$ for dark stripes aligned along the poloidal direction
and $B(u_n,v_n)=A(0,v-v_n)$ for dark stripes aligned along the toroidal direction.
After a particular initial seed is constructed in the isothermal coordinates,
it is converted to toroidal coordinates $(\phi,\theta)$ as per the
transformation~(\ref{eq:iso-tor}) or to Cartesian coordinates 
$(x,y)=(R\phi,r\theta)$ on the surface of the torus.
Note that, as per Eq.~(\ref{eq:3Dtorus}), $(X,Y,Z)$ describes the Cartesian 
coordinates for a point on the torus in 3D space while $(x,y)$ 
describes the Cartesian 2D coordinates on the {\em surface} of the torus.
Steady states are then found using Newton's method by discretizing space
using second order, central, finite differences (FDs) and separating real and
imaginary parts. We use a 2D grid of $N_\phi\times N_\theta$ mesh points
to discretize the wavefunction giving rise to a Newton matrix of size
$2N_\phi N_\theta\times 2N_\phi N_\theta$. In our numerics below we typically
use $N_\phi=N_\theta=490$.
Similarly, for the numerical stability results, we use the same FD discretization 
in space to cast the eigenvalue-eigenfunction problem (\ref{eq:mat}) as a standard
eigenvalue-eigenvector problem for the resulting stability matrix of size
$2N_\phi^2 N_\theta^2\times 2N_\phi^2 N_\theta^2$.
Finally, for the numerical integration of the full NLS (\ref{eq:nls}) we
use again the same FD discretization in space and a standard fourth order
Runge-Kutta (RK4) in time.

\begin{figure}
\includegraphics[width=1.0\columnwidth]{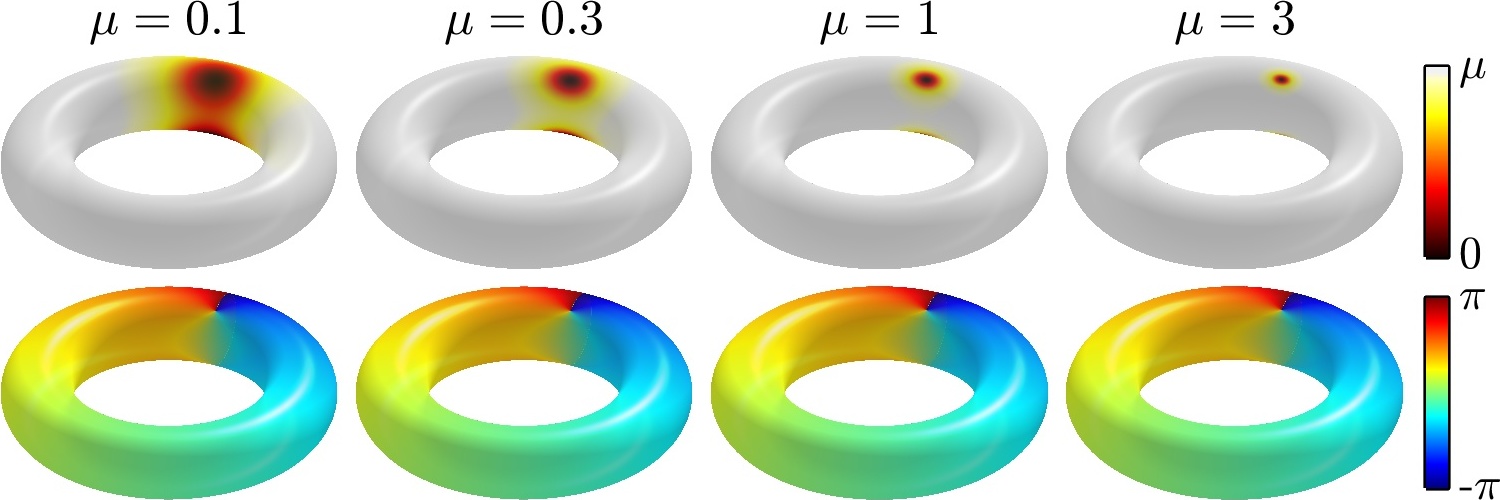}
\\[3.0ex]
\includegraphics[width=1.0\columnwidth]{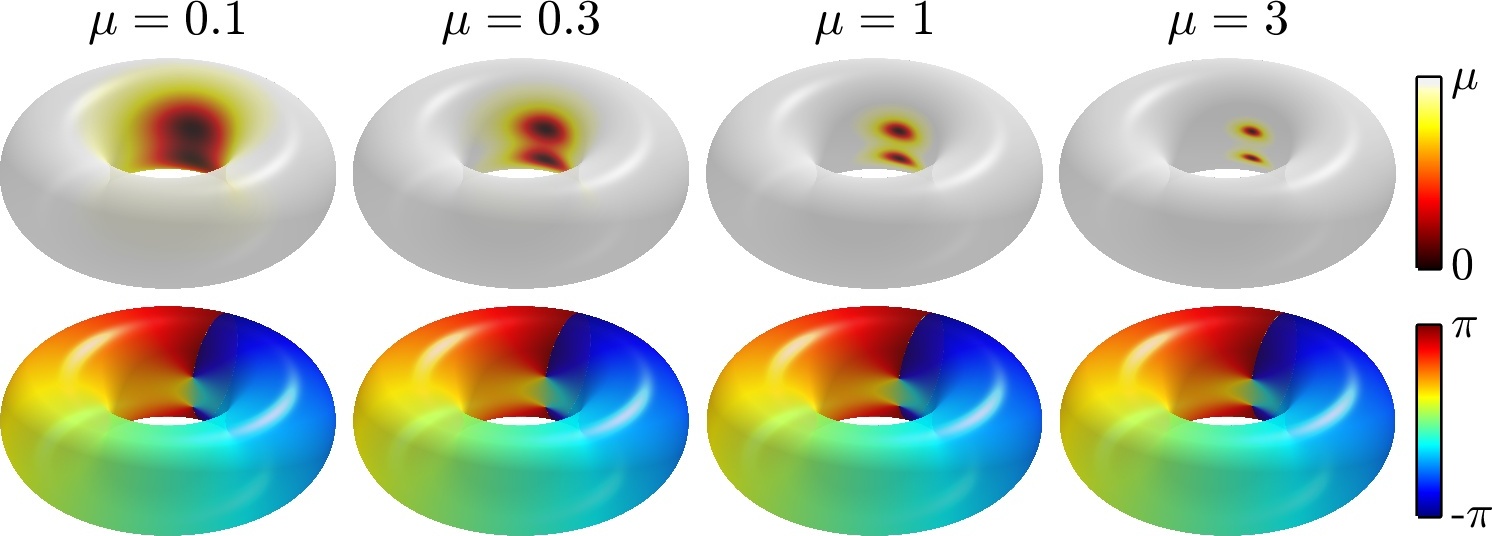}
\caption{(Color online)
Steady state solutions continued from the vertical dipole-in configuration.
The modulus squared (top) and the phase (bottom) of the solutions are
plotted on the surface of the torus.
The different solutions correspond to the values of $\mu$ indicated 
in the panels and $R=12$ while $\alpha=0.4$ for the top  group of panels 
and $\alpha=0.7$ for the bottom group of panels.
}
\label{fig:VIsamples}
\end{figure}

\begin{figure}
\includegraphics[width=1.0\columnwidth]{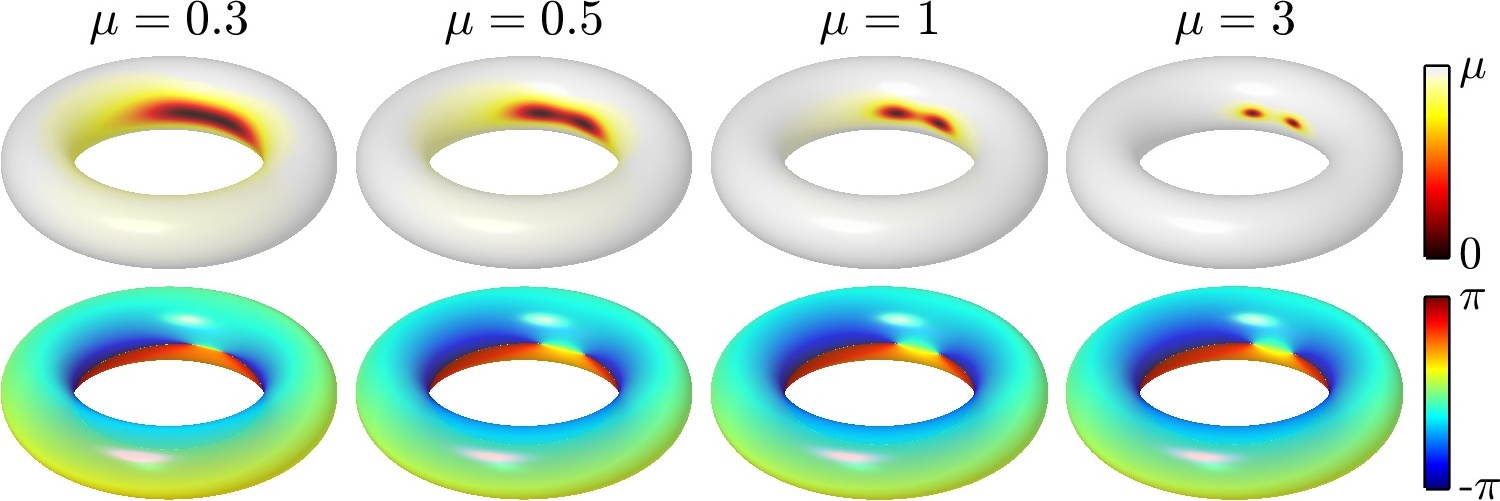}
\\[3.0ex]
\includegraphics[width=1.0\columnwidth]{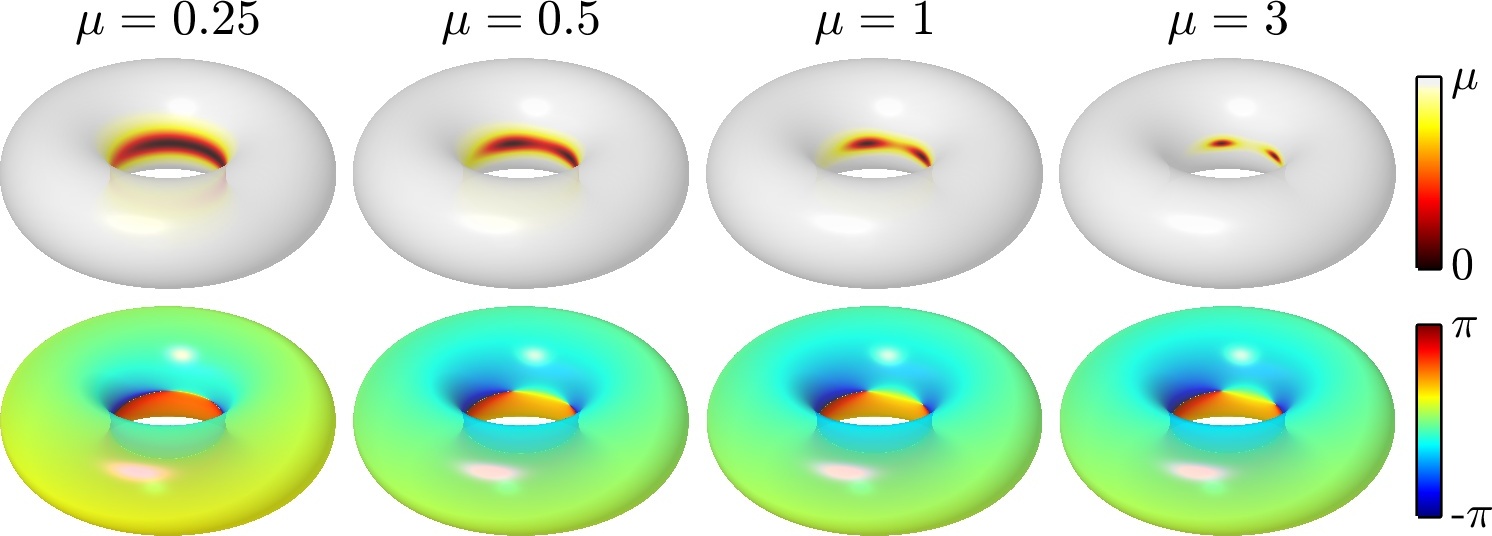}
\caption{(Color online)
Steady state solutions continued from the horizontal dipole-in configuration.
Same layout and parameters as in Fig.~\ref{fig:VIsamples}.
}
\label{fig:HIsamples}
\end{figure}

\begin{figure}
\includegraphics[width=1.0\columnwidth]{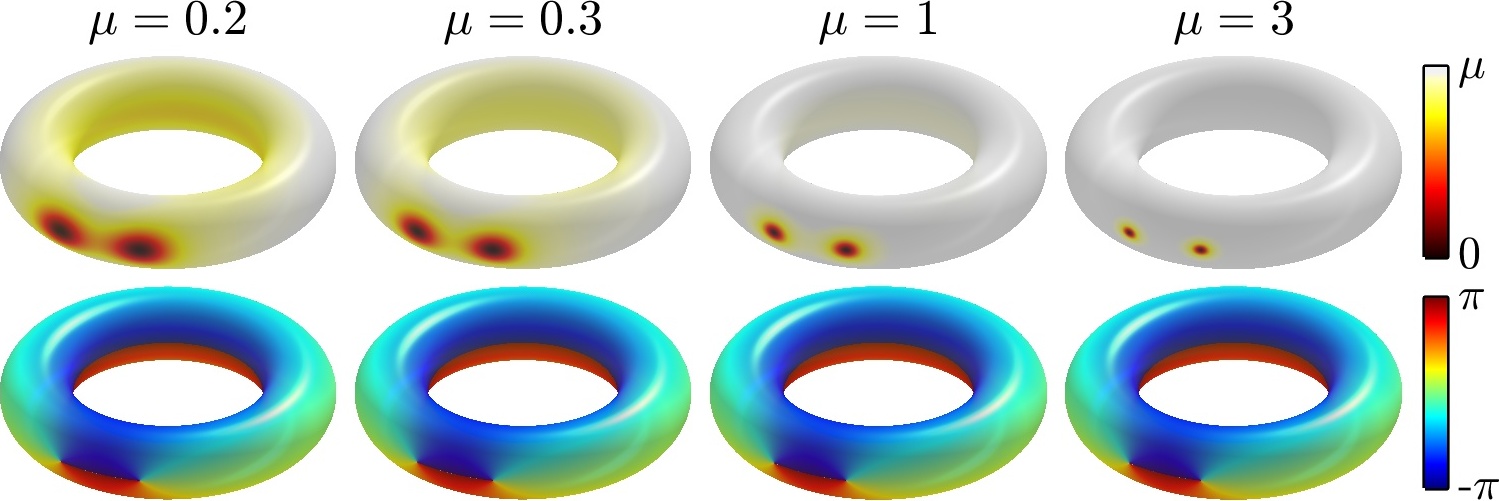}
\\[3.0ex]
\includegraphics[width=1.0\columnwidth]{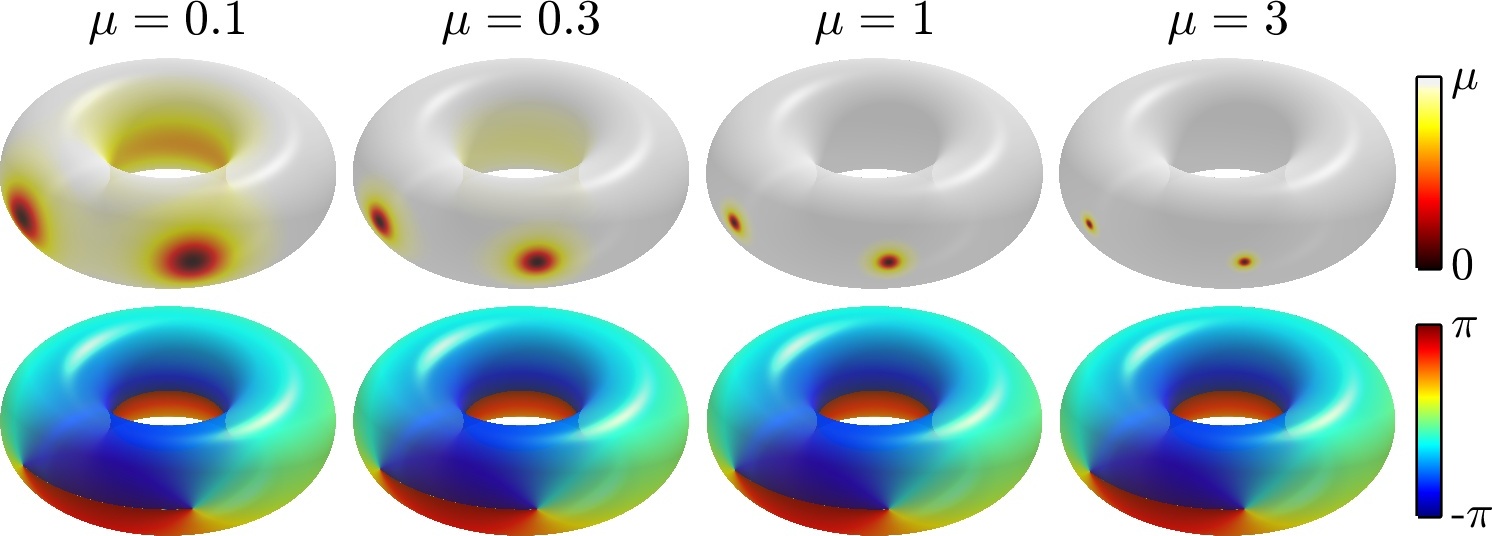}
\caption{(Color online)
Steady state solutions continued from the horizontal dipole-out configuration.
Same layout and parameters as in Fig.~\ref{fig:VIsamples}.
}
\label{fig:HOsamples}
\end{figure}

\begin{figure}
\includegraphics[width=1.0\columnwidth]{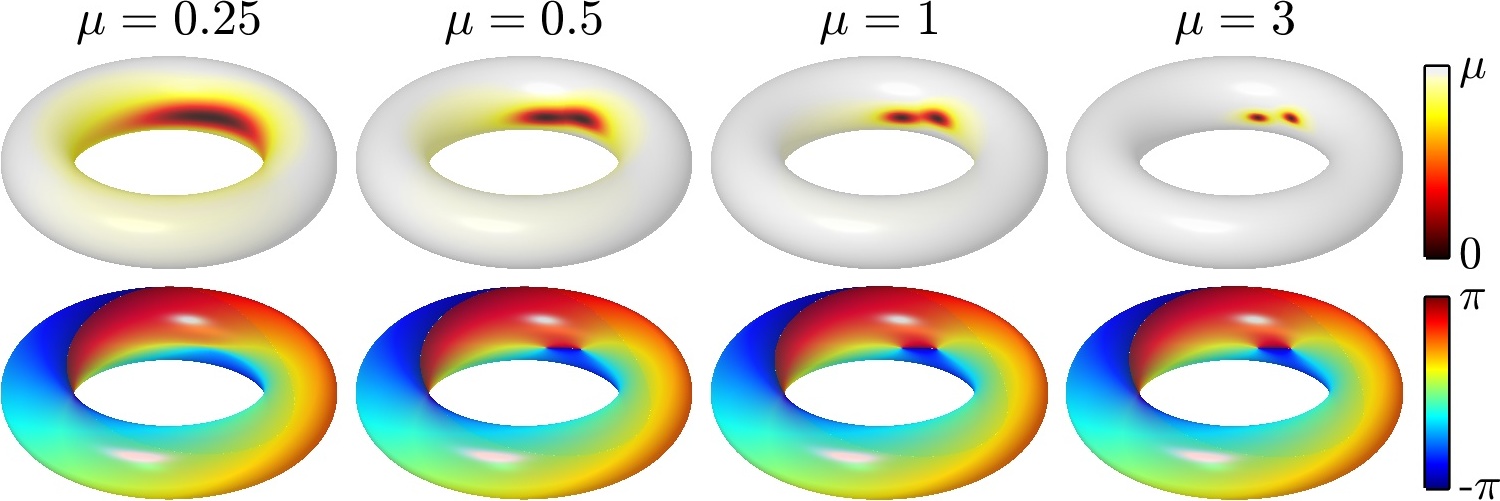}
\\[3.0ex]
\includegraphics[width=1.0\columnwidth]{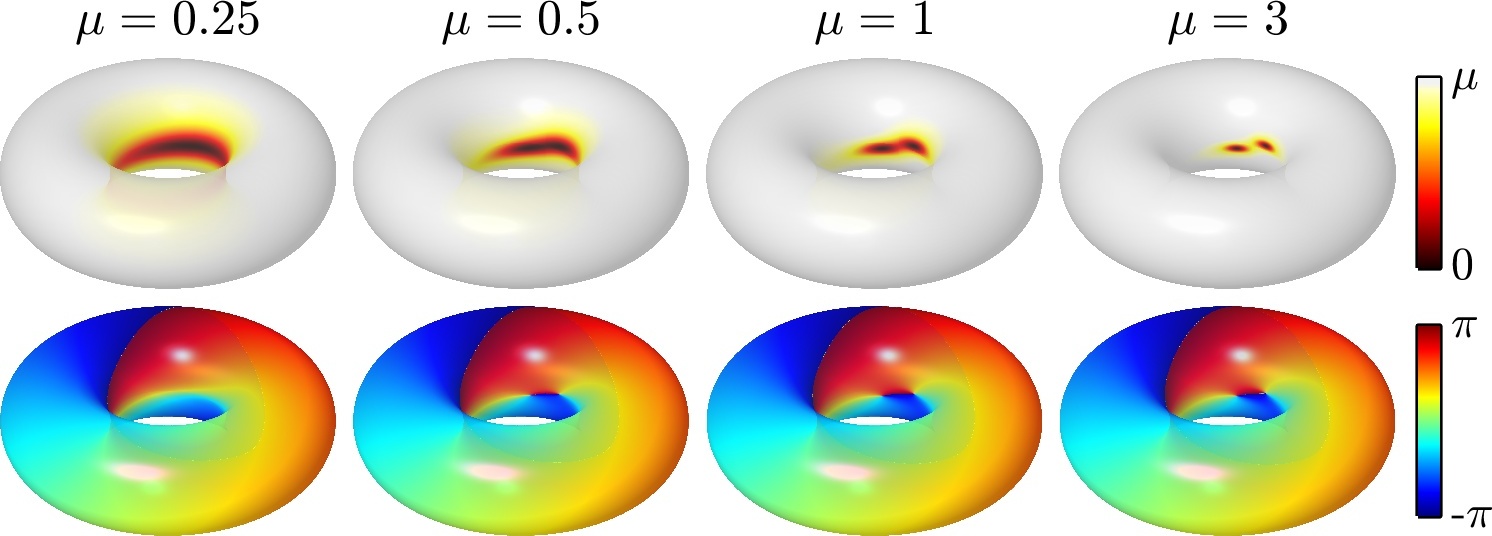}
\caption{(Color online)
Steady state solutions continued from the diagonal dipole configuration.
Same layout and parameters as in Fig.~\ref{fig:VIsamples}.
}
\label{fig:DIAGsamples}
\end{figure}

\begin{figure}[htb]
\includegraphics[width=0.95\columnwidth]{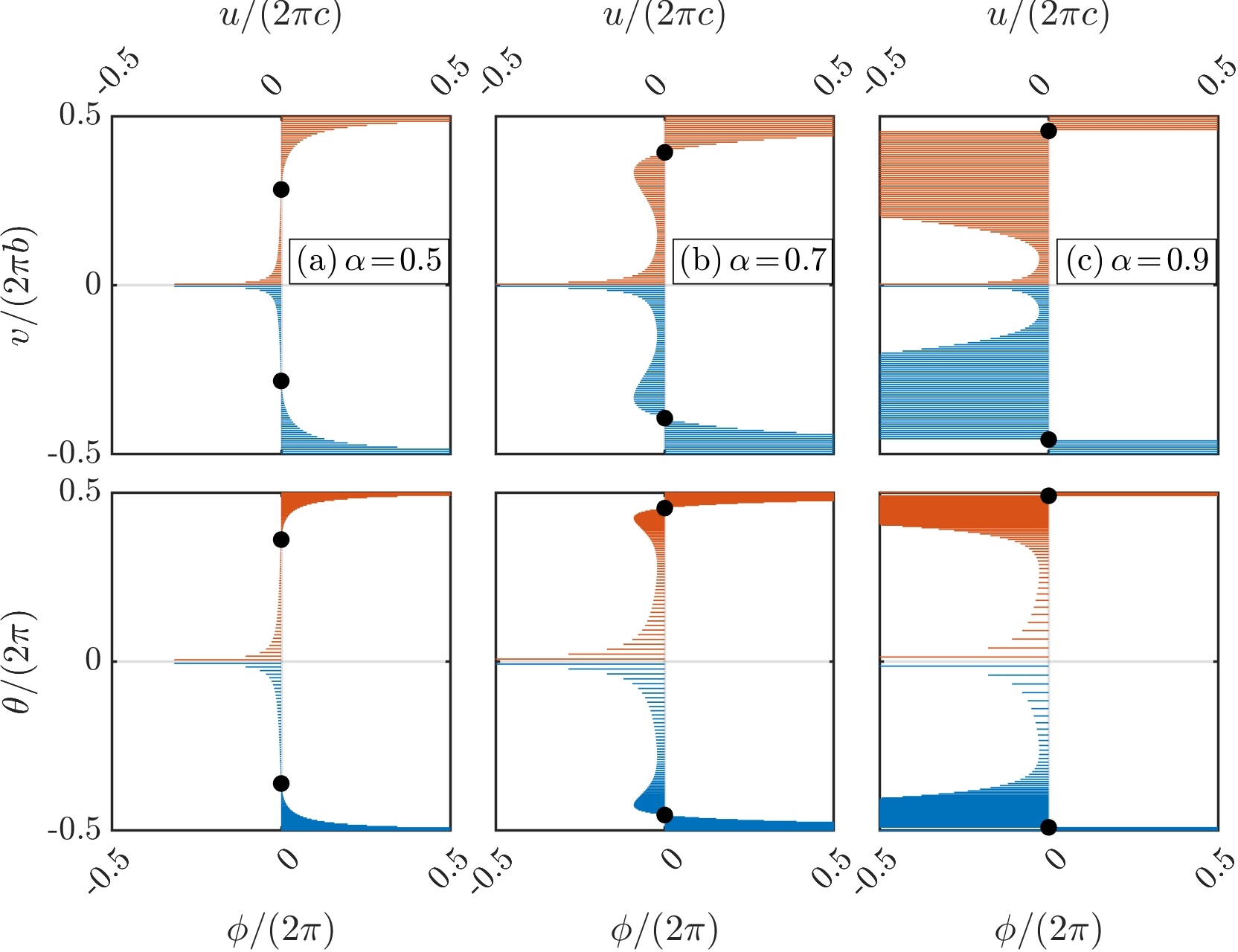}
\caption{(Color online)
Instantaneous velocities for vertical dipoles-in. 
Top and bottom row of panels depict, respectively, isothermal and toroidal 
coordinates and the different columns correspond to the indicated values for $\alpha$.
Blue and red curve correspond, respectively, to the velocities from 
the reduced ODE model for the plus and minus vortices. 
The different orbits such as those seen in Fig. \ref{fig:phase_space_hor_dip}  
are generated from initial conditions corresponding
to symmetric perturbations from the unstable stationary
vertical dipole-in positions (denoted by black circles). 
Notice that what changes between the columns is the value of the aspect ratio $\alpha$.
}
\label{fig:phase_space_vert_dip}
\end{figure}

\begin{figure}[htb]
\includegraphics[width=0.95\columnwidth]{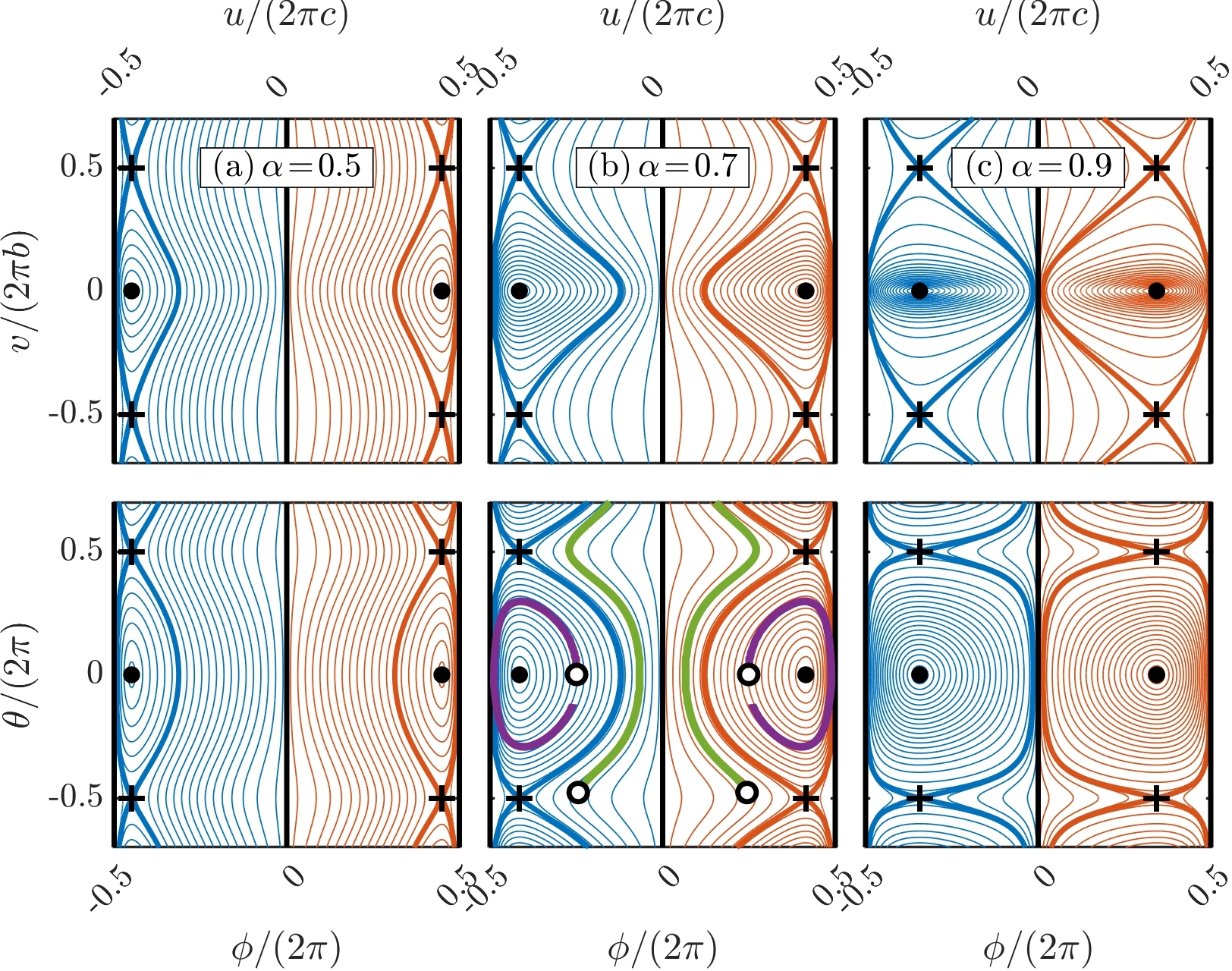}
\\
\center
~~
\includegraphics[width=0.40\columnwidth]{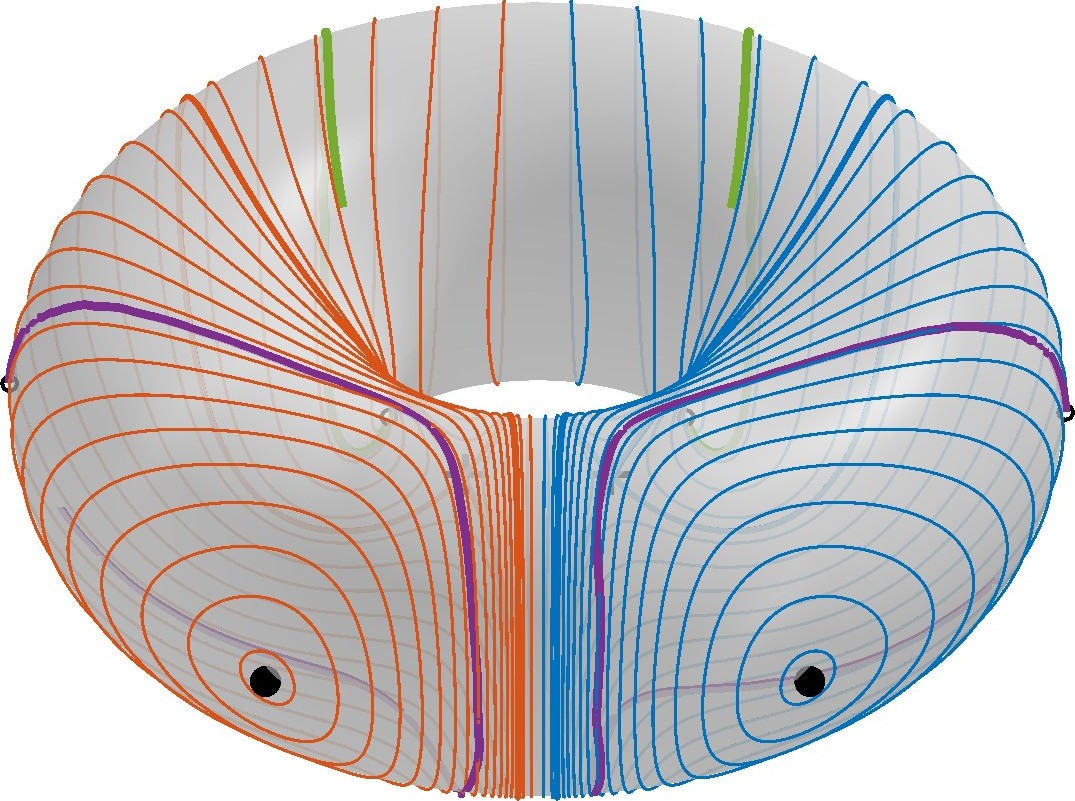}
\quad
\includegraphics[width=0.40\columnwidth]{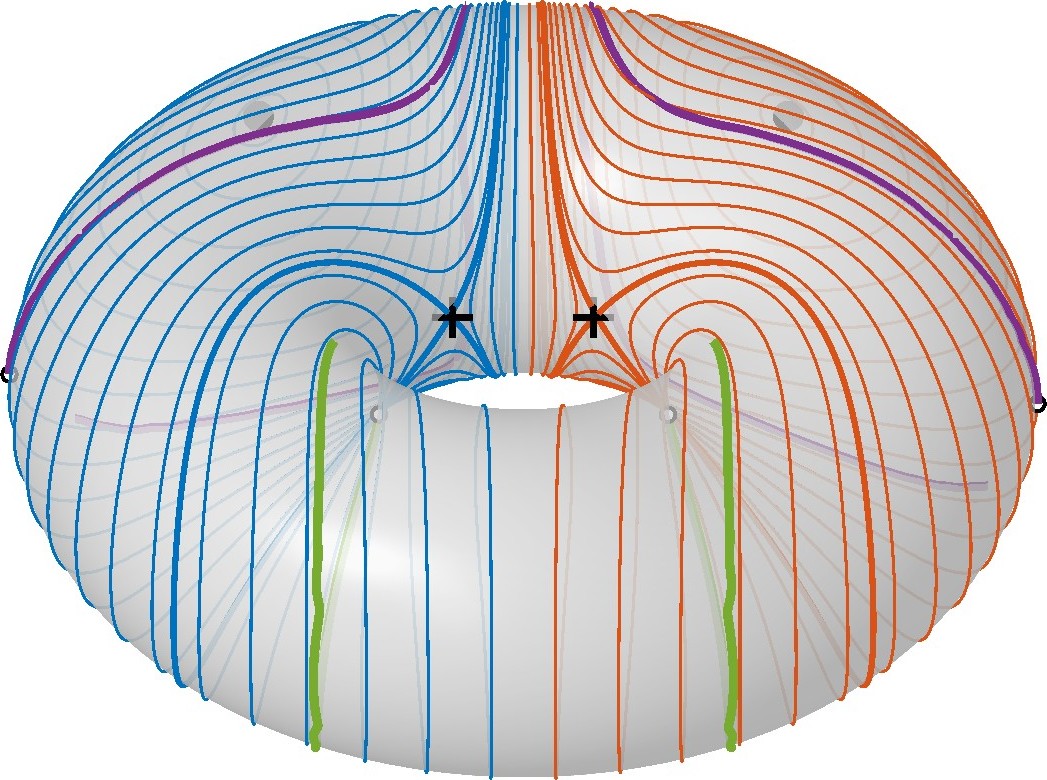}
\caption{(Color online)
Planar phase space for symmetric horizontal dipole configurations. 
Top and middle row of panels depict, respectively, isothermal and toroidal 
coordinates and the different columns correspond to the indicated values for $\alpha$.
The bottom row depicts the front (left subpanel) and back (right subpanel) 
views of the $\alpha=0.7$ case projected on the surface of the torus.
The different orbits are generated from initial conditions corresponding to symmetric 
perturbations from the (neutrally) stable horizontal dipole-out (black dots). 
The separatrices (thick curves) correspond to the stable 
and unstable manifolds of the horizontal dipole-in (black plus symbols).
The separatrices divide the phase space in areas containing oscillating 
orbits and rotating (wrapping poloidally) orbits.
For $\alpha=0.7$ we also depict the trajectories from full PDE
simulations for an oscillating orbit (thick purple line) and
a rotating one (thick green line) starting at the initial
positions depicted by the corresponding white dots that are located at the 
outer ($\theta = 0$) and inner ($\theta = \pi$) parts of the torus respectively.
}
\label{fig:phase_space_hor_dip}
\end{figure}

\begin{figure}[htb]
\includegraphics[width=0.85\columnwidth]{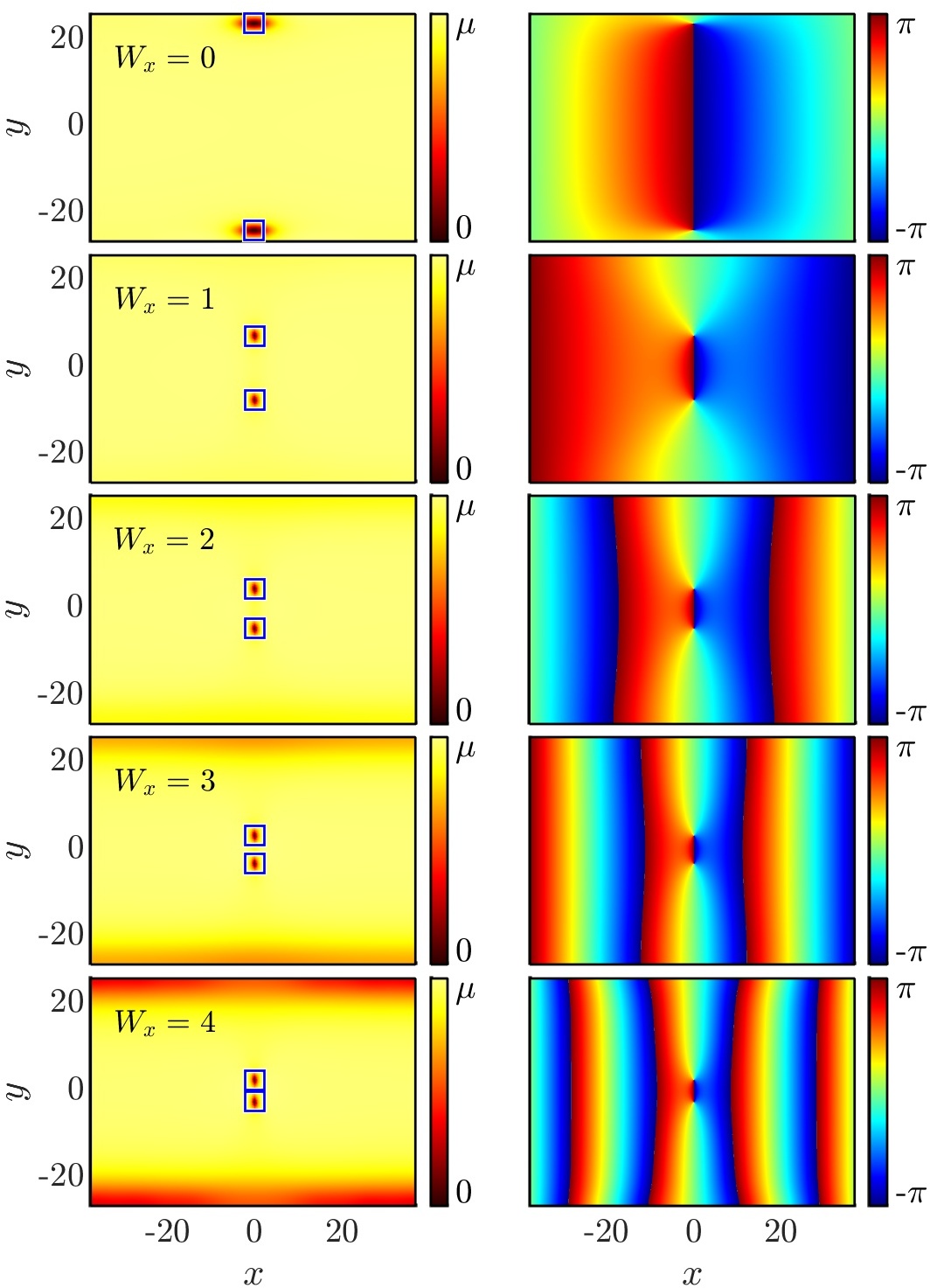}
\caption{(Color online)
Stationary vertical dipole solutions with increasing horizontal windings 
$W_x$ (and $W_y=0$) for $\alpha=0.7$, $\mu=1$, and $R=12$. 
The left and right columns depict, respectively, the
density and phase of the solutions in Cartesian coordinates.
The location of the vortices corresponding to the ODE model are overlayed 
to the density plot using blue squares.
The top row corresponds to the standard vertical dipole-in without
extra winding ($W_x=0$)
and each successive row corresponds to
a stationary vertical dipole solutions with increasing horizontal 
winding number $W_x$, with the vortices getting closer to compensate
for the additional speed in the opposite direction provided by the winding.
}
\label{fig:dip_ver_windings}
\end{figure}

\begin{figure}[htb]
\includegraphics[width=0.85\columnwidth]{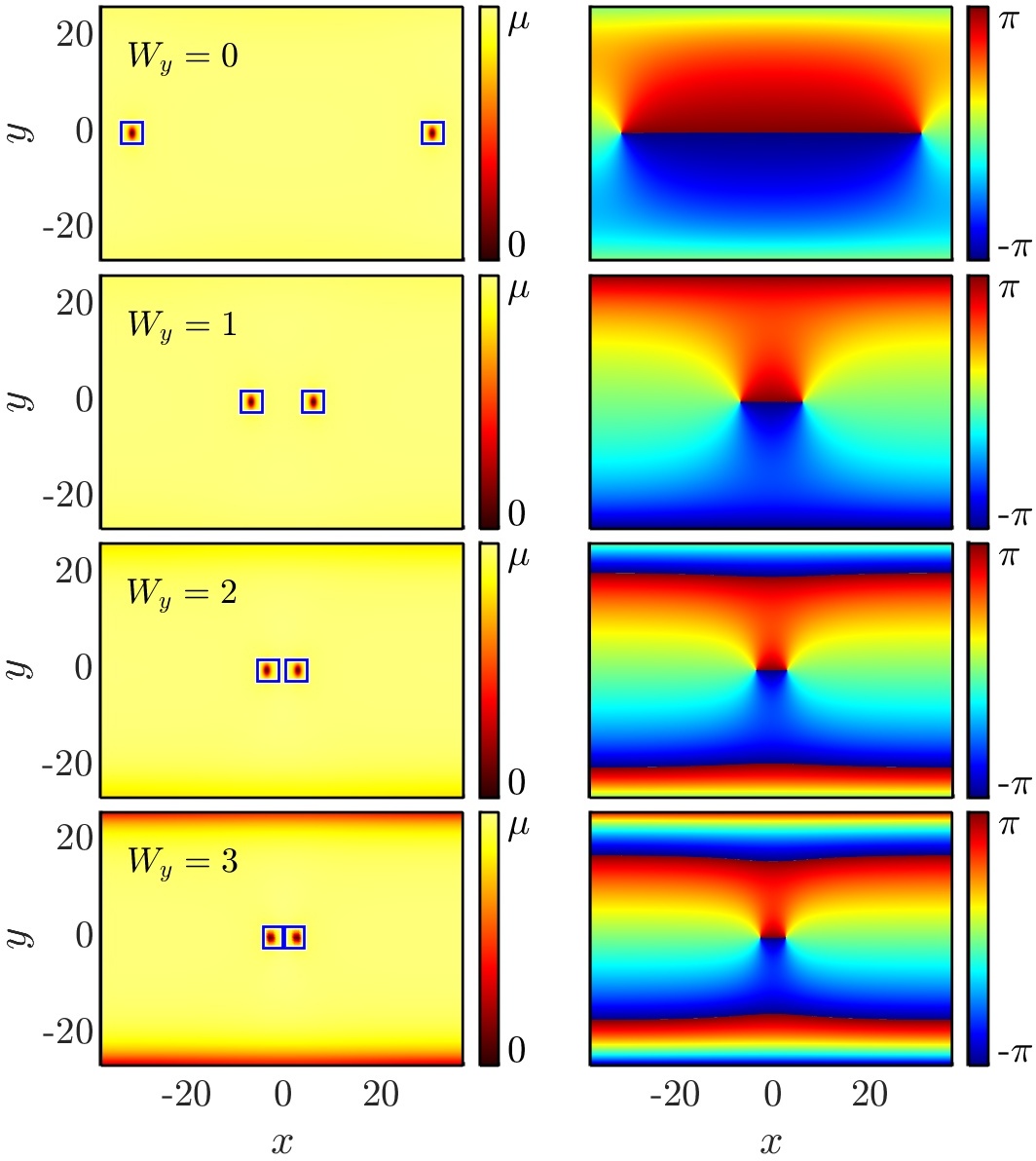}
\caption{(Color online)
Same as Fig.~\ref{fig:dip_ver_windings} but for stationary horizontal 
dipoles for $\alpha=0.7$, $\mu=1$, and $R=12$ with vertical windings
$W_y$ (and $W_x=0$).
}
\label{fig:dip_hor_windings}
\end{figure}

\subsection{Vortex Dipoles}
\label{sec:num:dip}

\subsubsection{Vortex Dipoles: steady states}
\label{sec:num:dip:steady}

Through the reduced ODE model, one can browse the entire
phase space of solutions for vortex dipoles given by a $+1$ vortex at 
location $(\phi_1,\theta_1)$ and a $-1$ vortex at $(\phi_2,\theta_2)$. Since 
the system is translationally invariant in the toroidal direction, the original
phase space $(\phi_1,\theta_1,\phi_2,\theta_2)$ can be reduced, without loss of
generality, to $(\bar\phi,\theta_1,-\bar\phi,\theta_2)$ by centering the solution
about the toroidal axis with $\bar\phi=(\phi_1+\phi_2)/2$.
A numerically exhaustive search for steady states in the three-dimensional reduced 
space $(\bar\phi,\theta_1,\theta_2)$, using a standard fixed point iteration 
method (nonlinear least squares with a Levenberg-Marquardt algorithm), is then 
straightforward and yields {\em four} different types of stationary dipoles. 
These correspond to:
\begin{itemize}
\item[$\circ$] {\bf Vertical dipole-in}. 
This solution corresponds to a vertically (poloidally) aligned dipole 
with $\phi_1=\phi_2$ and $\theta_1=-\theta_2$. We dub this solution to be `in'
as the value of $|\theta_1|=|\theta_2|$ is closer to $\theta=\pi$, the inner
part of the torus, than to the outer part with $\theta=0$. 
Figure~\ref{fig:VIsamples} depicts several steady state solutions continued
from the vertical dipole for $R=12$, a couple of values of $\alpha$, and 
for different values of $\mu$.
\item[$\circ$] {\bf Horizontal dipole-in}.
This solution corresponds to a horizontally (toroidally) aligned dipole 
with $\theta_1=\theta_2=\pi$,
i.e., on the inside of the torus;
see Fig.~\ref{fig:HIsamples}.
\item[$\circ$] {\bf Horizontal dipole-out}.
This solution corresponds to a horizontally (toroidally) aligned dipole 
with $\theta_1=\theta_2=0$,
i.e., on the outside of the torus;
see Fig.~\ref{fig:HOsamples}.
\item[$\circ$] {\bf Diagonal dipole}.
This solution corresponds to a diagonal dipole with 
$(\phi_2,\theta_2)=-(\phi_1,\theta_1)$ owing to a non-trivial balance 
of all the vortex velocity components;
see Fig.~\ref{fig:DIAGsamples}. This is, arguably, the least
intuitively expected among the different solutions.
\end{itemize}

At an intuitive level, one can argue that the main phenomenology
involves a combination of different factors.
On the one hand, a well-known fact stemming from their nonlinear, 
phase-induced interaction is that two vortices in Euclidean
space will travel parallel to each other (in a direction perpendicular
to their line of sight). The curvature arising from
the toroidal geometry leads to that feature being relevant now in the
isothermal coordinates (where locally the metric is conformal to the
Euclidean one). On the other hand, the topology of the torus and its periodic
boundary conditions come into play and effectively create an equal
and opposite velocity at these suitably selected distances, creating
the potential for the steady states that we consider herein.

It is possible to analyze further the statics and dynamics of the vertical
and horizontal dipoles as, per the reduced ODE model, any initial
condition corresponding to (i) a symmetric ($v_1=-v_2$) vertical dipole
or to (ii) a horizontal dipole will always remain (i) a symmetric 
vertical dipole or (ii) a horizontal dipole.
Figures \ref{fig:phase_space_vert_dip} and \ref{fig:phase_space_hor_dip}
show, respectively, the instantaneous velocities for vertical dipoles-in and
the planar phase space $(u,v)$ [or $(\phi,\theta)$] for horizontal dipoles.
A full explanation of the results for vertical dipoles presented in 
Fig.~\ref{fig:phase_space_hor_dip} is given below in Sec.~\ref{sec:num:dip:dyn}.

With regards to the vertical steady state dipole, 
this solution exists as there is no poloidal contribution to the vortex
velocities (velocity contribution from curvature effects is purely toroidal 
and vortex-vortex velocity contribution is purely toroidal) and the
toroidal contribution from curvature balances the vortex-vortex contribution
(also toroidal). 
In fact, as Fig.~\ref{fig:phase_space_vert_dip} suggests, for each 
value of $\alpha$ there seems to be a single vertical dipole distance 
that balances all velocity contributions leading to a steady state. 
The figure also suggests that the steady state vertical dipole is 
always {\em unstable} as, e.g., perturbations along the poloidal direction
will naturally result in
(i) a constant toroidal velocity of the dipole if the vortex perturbations 
are equal and opposite in the poloidal direction, and, for generic perturbations,
(ii) a toroidal velocity imbalance will start deviating exponentially from 
the stationary state.
This instability of the vertical dipole-in will be revisited,
for both PDE and ODE models, in Sec.~\ref{sec:num:dip:stab}.

On the other hand, as depicted in Fig.~\ref{fig:phase_space_hor_dip}, the
planar phase space for horizontal dipoles is much richer than the one
corresponding to vertical dipoles. 
The figure suggests that there exist two horizontal dipoles as
we described above: one at $\theta=0$ (horizontal dipole-out, denoted
by black dots in the figure) and 
one at $\theta=\pi$ (horizontal dipole-in, denoted by black plus symbols).
Furthermore, the figure suggests that the horizontal dipole-out
is (neutrally) stable as a center within the relevant phase portrait,
while the horizontal dipole-in is unstable (i.e., a saddle point).
As for the vertical dipoles, the stability for horizontal dipoles will be 
covered, for both PDE and ODE models, in Sec.~\ref{sec:num:dip:stab}.

By following the construction of steady state vertical dipoles as per 
Fig.~\ref{fig:phase_space_vert_dip}, it is apparent that there is no vertical 
dipole that could be dubbed `out'. This fact can be rationalized by noting that
the velocity contribution from curvature close to $\theta=0$ (i.e., on the outer 
part of the torus) and the velocity contribution from the vortex-vortex interaction 
are in the {\em same direction}. 
Nonetheless, let us note that it is possible, in the NLS model, to add
extra phase windings along the toroidal and poloidal directions provided
that one does respect the periodicity of the domain. In fact, any NLS
configuration $\psi(\phi,\theta)$ can always
be multiplied by a phase term with $W_x$ bearing an extra $2\pi$ winding in the
toroidal direction and $W_y$ an extra $2\pi$ winding in the poloidal direction
without violating the periodicity of the domain.
Thus, inspired by the vertical and horizontal dipoles described above,
one can take each of these solutions and multiply it by a phase
terms as follows:
\begin{equation}
\psi(\phi,\theta) ~\rightarrow ~
\psi(\phi,\theta)\cdot e^{iW_x \phi}\cdot e^{iW_y \theta},
\end{equation}
where the windings $W_x$ and $W_y$ are integers.
In a sense these structures bear two sets of topological charges,
with one stemming from the charge of the vortex constituents,
while the second arises through the potential windings along the
toroidal or poloidal (or both) directions around the torus.
Using this idea, we took the vertical-in and horizontal-out steady state dipoles
and progressively applied, respectively, horizontal and vertical windings
in tandem with the fixed point iteration scheme (Newton) to obtain families
of dipoles with higher windings. We showcase examples of higher-winding
vertical and horizontal dipoles-in Figs.~\ref{fig:dip_ver_windings} and
\ref{fig:dip_hor_windings}, respectively.
It is interesting to note that the vertical dipole steady state with 
$W_x \geq 1$ gives rise to vertical dipoles that could be dubbed `out'
as, in order to balance the extra vertical winding 
the vortices have to get close to each other around $\theta=0$.

It is possible to leverage the reduced equations of motion to include the
effects of vertical and horizontal windings. This relies on assuming
that both toroidal and poloidal contributions to the phase windings
in isothermal coordinates are accounted for by {\em linear} phase gradients.
Under this assumption, we incorporate linear phase windings that gain
$2\pi W_x$ and $2\pi W_y$ along, respectively, the horizontal
and vertical directions. These windings are captured by adding corresponding
linear terms in $\Omega_n$ of Eq.~(\ref{eq:ODEvel}) as follows:
\begin{equation}
\label{eq:ODE_windings}
\Omega_n \rightarrow \Omega_n +
\frac{1}{\Lambda(v_n)}\left[ i \frac{W_x}{c} + \frac{W_y}{r}\right].
\end{equation}
The corresponding fixed points obtained from this extended reduced model 
with vertical and horizontal windings are depicted by the blue squares in 
the different panels of, respectively, Figs.~\ref{fig:dip_ver_windings} 
and \ref{fig:dip_hor_windings}.
As evidenced from these figures, this extended reduced ODE accurately
predicts the stationary locations of vortex dipole configurations including
vertical and horizontal windings.
Furthermore, as we will see in the next section, this extended reduced ODE
will also accurately describe the stability properties of these stationary
dipole configurations.

\begin{figure}[htb]
\includegraphics[width=0.95\columnwidth]{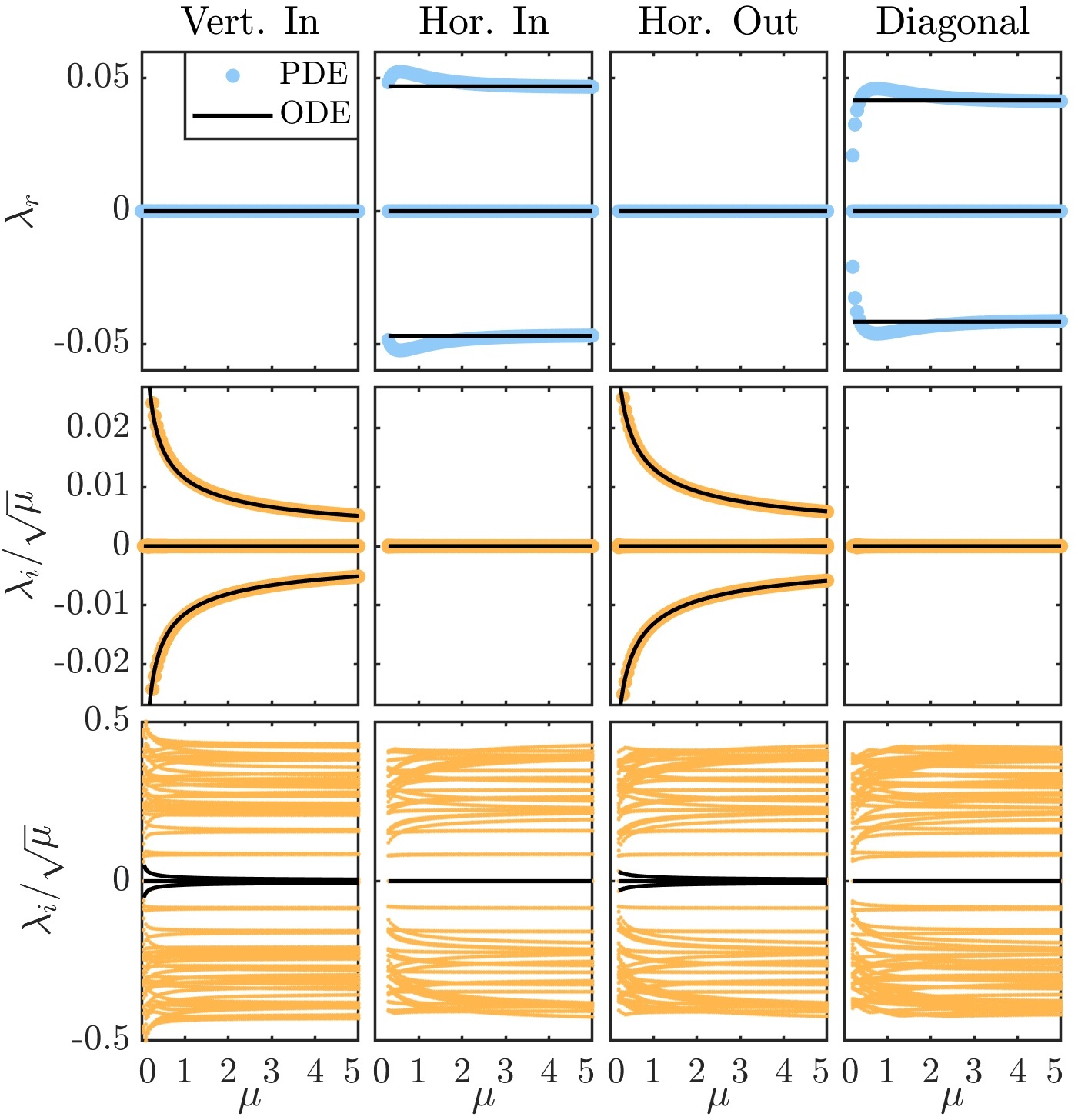}
\caption{(Color online)
Convergence of the stability spectrum for dipole configurations as
the chemical potential $\mu$ is increased for $\alpha=0.4$ and $R=12$.
The columns correspond, from left to right, to the vertical
dipole-in, horizontal dipole-in, horizontal dipole-out, and diagonal
dipole configurations.
The top row of panels depicts the real part of the linearization
eigenvalues $\lambda_r$ where the large (light blue) dots correspond to
the full BdG spectra computation of the PDE while the black solid
line depicts the corresponding results from the reduced ODE model.
Note that $\lambda_r>0$ is associated with an {\em unstable} solution.
The bottom two rows depict the imaginary part of the linearization
eigenvalues $\lambda_i$ (normalized by $\sqrt{\mu}$) where, again, 
the large (orange) dots correspond to the PDE while the black solid 
line corresponds to the ODE. 
The middle row of panels is a zoomed in version of the bottom row.
}
\label{fig:spectra_mu_alpha04}
\end{figure}

\begin{figure*}[htb]
\includegraphics[width=1.94\columnwidth]{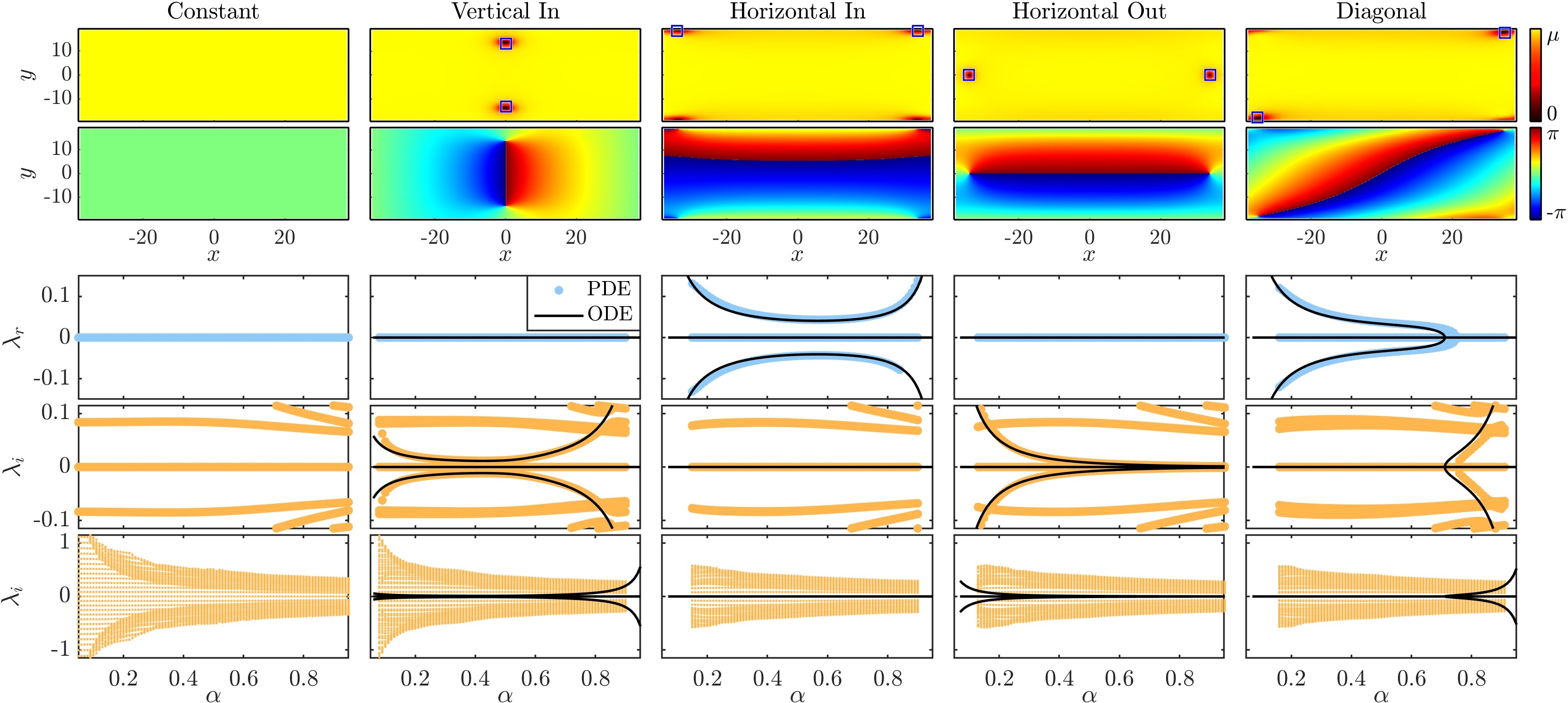}
\caption{(Color online)
Stability for dipole configurations as $\alpha$ is varied for $\mu=1$
and $R=12$.
The columns correspond, from left to right, to the constant, vertical
dipole-in, horizontal dipole-in, horizontal dipole-out, and diagonal
dipole configurations.
The top two rows depict, respectively, the density and phase 
of the steady state PDE solution for $\alpha=0.5$ in Cartesian coordinates. 
The location of the vortices corresponding to the ODE model are overlayed 
to the density plot using blue squares.
The third row of panels depicts the real part of the linearization
eigenvalues $\lambda_r$ where the large (light blue) dots correspond to
the full BdG spectra computation of the PDE while the black solid
line depicts the corresponding results from the ODE model.
The bottom two rows depict the imaginary part of the stability
eigenvalues $\lambda_i$ where, again, the large (orange) dots 
correspond to the PDE while the black solid line corresponds
to the ODE. 
The fourth row of panels is a zoomed in version of the bottom row.
}
\label{fig:spectra_alpha_mu1}
\end{figure*}

\begin{figure*}[htb]
\includegraphics[width=1.94\columnwidth]{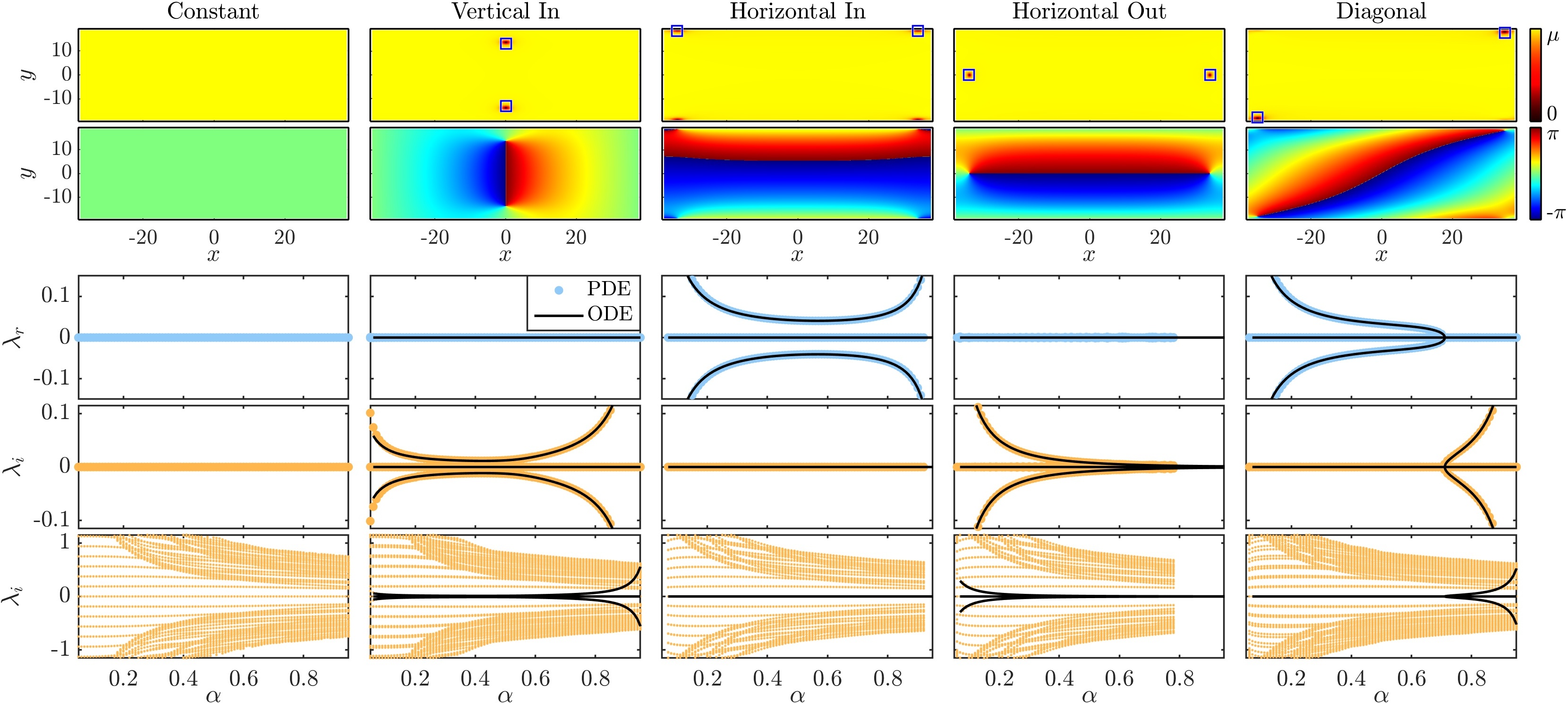}
\caption{(Color online)
Same as Fig.~\ref{fig:spectra_alpha_mu1} but for $\mu=5$.
Notice the better match between the PDE and ODE results as the
latter is obtained for the point-vortex model in the large $\mu$ limit.
}
\label{fig:spectra_alpha_mu5}
\end{figure*}

\begin{figure}[htb]
\includegraphics[width=0.90\columnwidth]{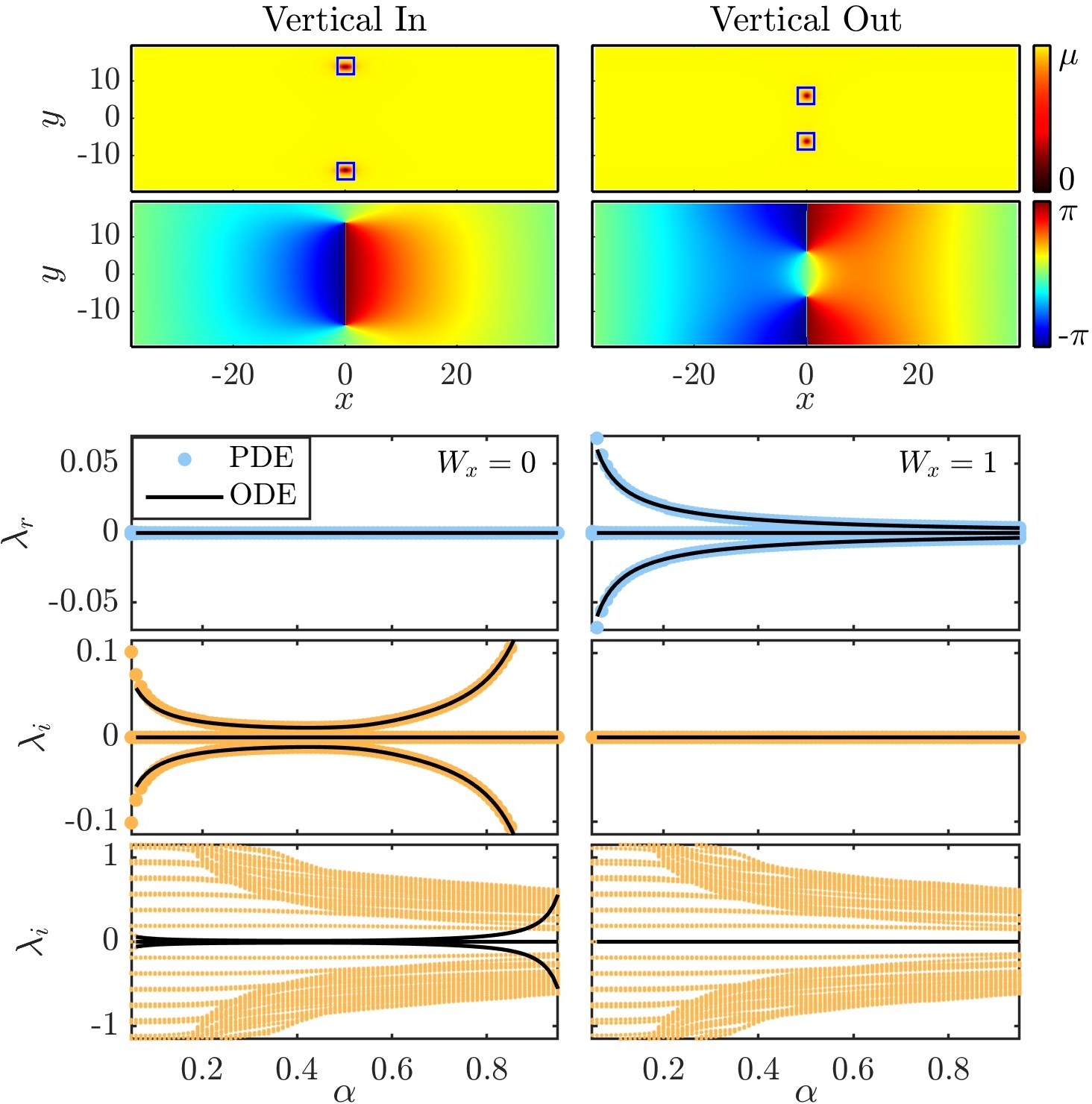}
\caption{(Color online)
Stability spectra vs.~$\alpha$ for the stationary vertical dipoles 
without any winding ($W_x=0$; left panels) and with one horizontal 
winding ($W_x=1$; right panels) for $R=12$ and $\mu=5$;
see top two rows of Fig.~\ref{fig:dip_ver_windings}.
Same layout and parameters as in Fig.~\ref{fig:spectra_alpha_mu5}.
Note that the addition of a horizontal winding destabilizes the
vertical dipole.
}
\label{fig:spectra_vertInOut}
\end{figure}

\subsubsection{Vortex dipoles: stability}
\label{sec:num:dip:stab}

Equipped with the four steady state dipole solutions described in the
previous section, let us now study their stability properties at the
full NLS and reduced ODE levels.
Figure~\ref{fig:spectra_mu_alpha04} depicts the stability spectra
for the four dipole solutions in both the full NLS (large colored dots)
and the reduced ODE model (black curves) as $\mu$ is varied.
We note that the stability eigenvalues $\lambda=\lambda_r+i \lambda_i$
determine the stability of the corresponding solutions as follows:
(i) $\lambda_r=0$ corresponds to a (neutrally) stable solution,
(ii) $\lambda_r\not=0$ and $\lambda_i=0$ corresponds to an exponential instability,
and
(iii) $\lambda_r\not=0$ and $\lambda_i\not=0$ corresponds to an oscillatory instability.
As Fig.~\ref{fig:spectra_mu_alpha04} indicates, for the parameters used
(namely $\alpha=0.4$ and $R=12$), the vertical-out and horizontal-out dipoles 
are {\em stable} while the horizontal-in and diagonal dipoles are {\em unstable}.
Importantly, the figure also evidences the striking match of the reduced
ODE model and the PDE findings, with the former very accurately
capturing the eigenvalues associated with the motion of the vortices.
Recall that the reduced ODE model is predicated on the condition of
the vortices representing a point particle.
This is certainly true in the large $\mu$ limit where the size of the vortex cores 
(tantamount to the healing length of $\propto 1/\sqrt{2 \mu}$) tends to zero. 
However, even for relatively small values of $\mu$, the particle model
prediction for the eigenvalues remains remarkably accurate.
Indeed, even moderate values of $\mu\gtrsim 3$ converge such that the relative
error for $\lambda_r$ and $\lambda_i$ are always less than $1\%$.
However, it is important to mention that configurations bearing vortices 
that are closer than a few times their width will not
be accurately captured by the reduced ODE. In fact, extreme cases could
lead to the annihilation of opposite charged vortices in the full NLS,
while such a scenario does not arise in the reduced ODE model that considers
the vortices as point-particles (with zero width).

Let us now study the bifurcation scenarios when the aspect ratio $\alpha$ of the
torus is varied. This non-trivial effect changes in a nonlinear fashion the 
relative size of the vortex-vortex contributions and the curvature effects and,
thus, one could expect interesting bifurcations.
Figures~\ref{fig:spectra_alpha_mu1} and \ref{fig:spectra_alpha_mu5} depict the
stability eigenvalues (alongside typical solutions) for the constant background
state and the four possible dipole configurations (with $W_x=W_y=0$; namely
without any extra windings) respectively, for $\mu=1$ and $\mu=5$.
The spectrum for the constant background is supplied in the figures so that
one is able to identify the eigenvalues that come from the actual vortices
and which ones stem from the background where they are embedded.
Naturally, the ODE model is only able to capture the former set of eigenvalues
originating exclusively from the relative motion of the vortices.
As before, we note that the reduced ODE model reproduces remarkably well
the relevant eigenvalues. In particular, for $\mu=5$ the
match between the NLS and reduced ODE spectra is striking, although it
should be noted that the relevant match is fairly reasonable even for $\mu=1$. 
In fact, the reduced ODE is able to perfectly capture (qualitatively and
quantitatively) the bifurcation suffered by the diagonal dipole where 
it is rendered {\em stable} for $\alpha\gtrsim 0.71$ (for $\mu=5$).
For the other solutions, as it was shown in Fig.~\ref{fig:spectra_mu_alpha04},
the spectra for different torus aspect ratio $\alpha$ of 
Figs.~\ref{fig:spectra_alpha_mu1} and \ref{fig:spectra_alpha_mu5} tend
to indicate that the vertical-in and horizontal-out dipoles are {\em stable} 
while the horizontal-in and diagonal dipoles are {\em unstable}.
In each case, the effective particle equations bear a vanishing
eigenvalue associated with the neutrality of the relevant solutions
against shifts along the toroidal direction $\phi$. It is thus only
the remaining pair of eigenvalues and the pertinent ``internal mode''
of the dipole dynamical motion that is responsible for the stability
(in the case of the vertical-in and horizontal-out dipoles) and for
the instability (for the remaining horizontal-in and diagonal cases).

To complement the stability results we include in 
Fig.~\ref{fig:spectra_vertInOut} the stability spectra of the vertical
dipole solution alongside its corresponding dipole with a winding $W_x=1$. 
As the figure suggests, adding a winding completely changes the
stability picture by destabilizing 
the vertical dipole solution. This is in line with the general expectation
that higher winding wave patterns are less likely to be dynamically
robust than lower winding ones.
Furthermore, we again obtain a remarkable agreement of the stability spectra
results between the full NLS and, in this case, the extended reduced ODE 
model~(\ref{eq:ODE_windings}) including vertical and horizontal windings.
Further studies on solutions bearing windings, including mixed combinations of
vertical {\em and} horizontal windings, at the NLS and
ODE levels are outside the scope of the present work and are thus left
for future explorations.

\begin{figure}[t]
\includegraphics[width=4.1cm,height=2.5cm]{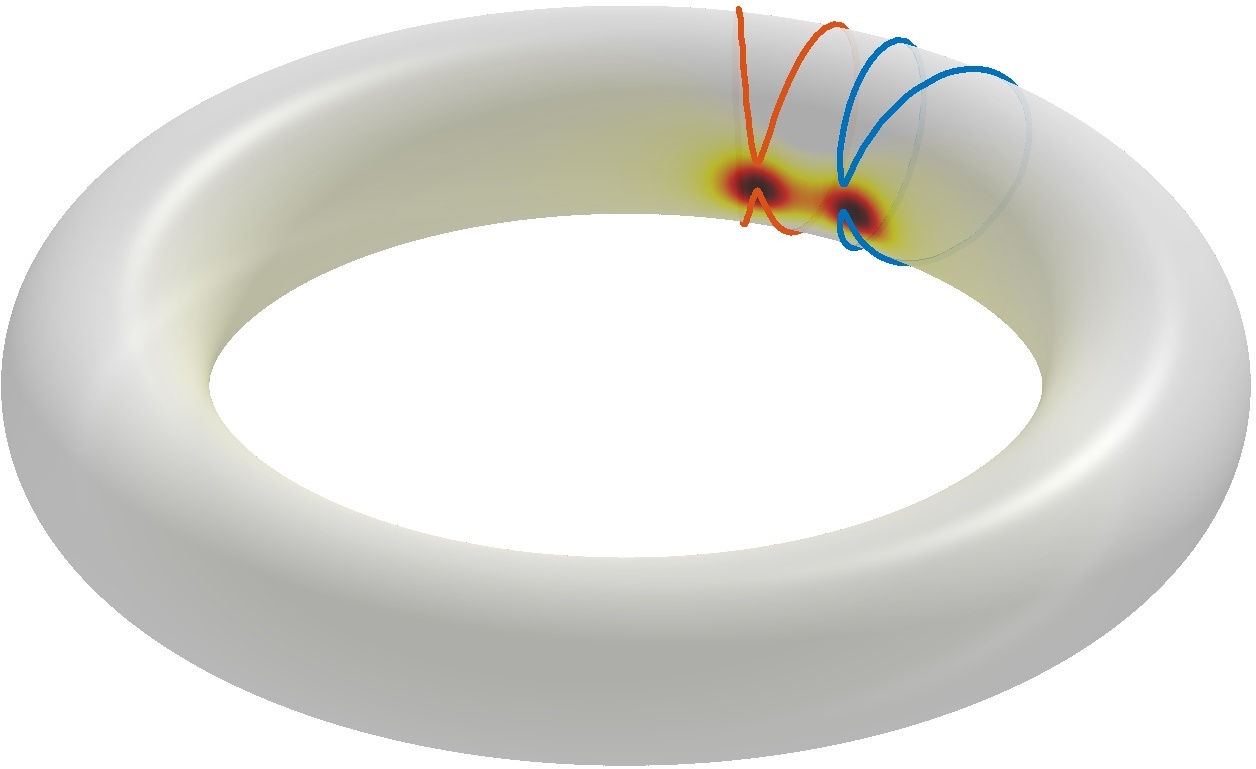}
~
\includegraphics[width=4.1cm,height=2.5cm]{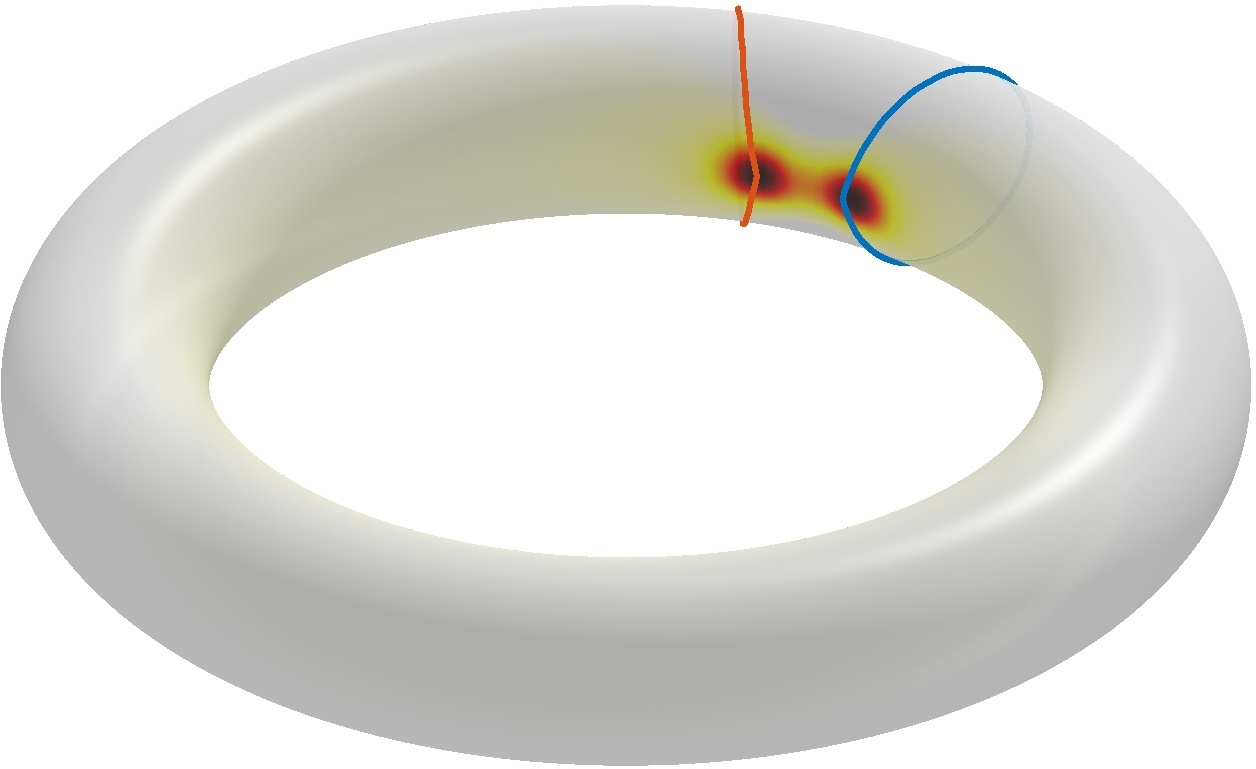}
\\[2.0ex]
\includegraphics[width=4.1cm,height=2.5cm]{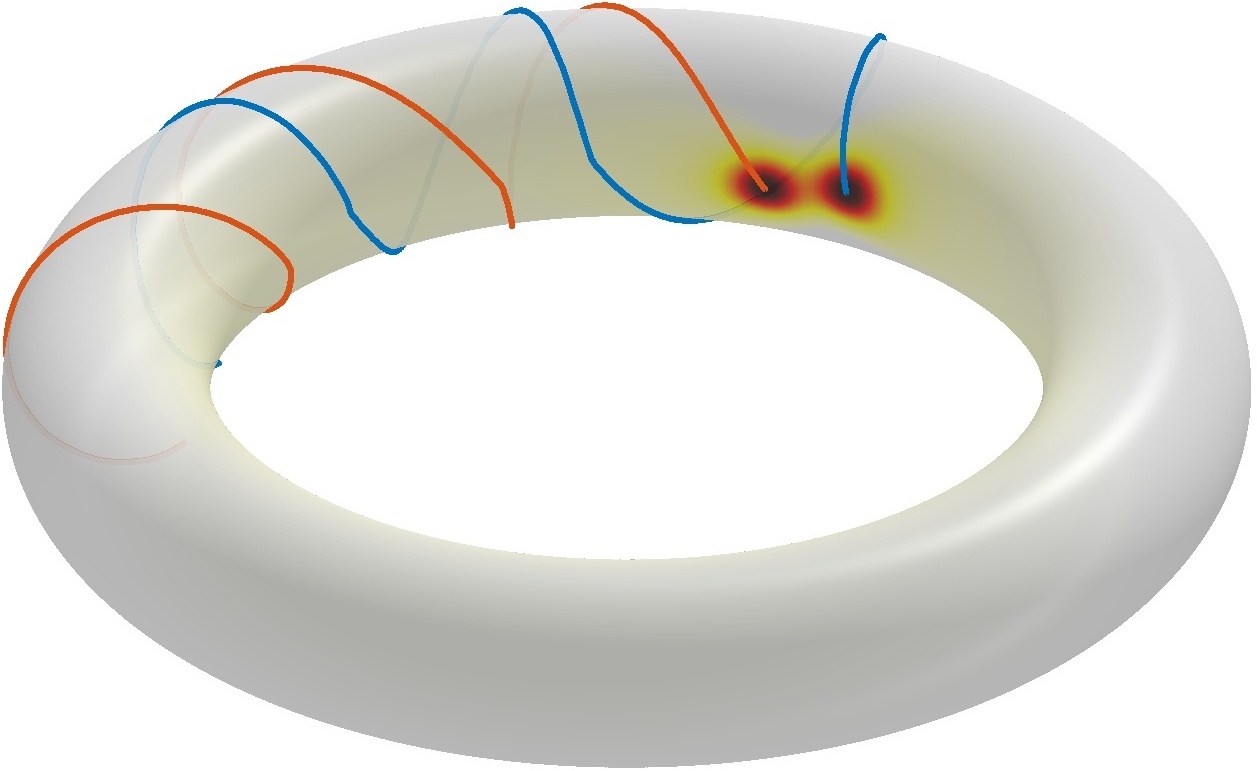}
\caption{(Color online)
Dynamics ensuing from the destabilization of unstable dipole configurations.
The top panels depict one period of oscillating (left) and
rotating (right) dipole-in orbits close to the separatrix.
The oscillating and rotating orbits were obtained by slightly shifting 
the vortices in the poloidal and toroidal directions, respectively.
The bottom panel depicts the destabilization of the diagonal dipole.
All cases correspond to $\mu=5$, $R=12$, and $\alpha=0.2$
The colored surface depicts the initial density and the overlaid curves
correspond to the trajectory traces from the negative (red) and positive 
(blue) vortices.
Please see 
{\tt HorIn-OSC-R12-alpha02-mu5.gif}, 
{\tt HorIn-LIB-R12-alpha02-mu5.gif}, and 
{\tt Diag-R12-alpha02-mu5.gif} 
in the Supplemental Material for respective
movies depicting the evolution of the density and phase.
}
\label{fig:dip_dynamics}
\end{figure}

\subsubsection{Vortex dipoles: dynamics}
\label{sec:num:dip:dyn}

In this section we present some results pertaining the dynamics of
vortex dipole configurations.
For instance, Fig.~\ref{fig:phase_space_hor_dip} depicts the dynamics for
horizontal dipoles from the reduced ODE model. 
Interestingly, the stable and unstable manifolds of the unstable (saddle) 
horizontal dipole-in coincide in a homoclinic orbit and define a 
separatrix between oscillating (librating) solutions around the horizontal 
dipole-out and rotating solutions that wrap along the poloidal direction.
In the case of $\alpha=0.7$ we also include two NLS orbits obtained
from initial positions as indicated in the figure and a phase profile
given, as before, by $\Phi(w)={\rm Im}(\sum_n q_n F(w,w_n))$.
The NLS orbits were extracted by doing a local fit of the extrema
of the vorticity, defined as the curl of the velocity, with the latter
defined as is standard in NLS settings~\cite{siambook}, i.e.,
\begin{eqnarray}
  \mathcal{V}=-\frac{i}{2} \frac{\psi^* \nabla \psi - \psi \nabla
  \psi^*}{|\psi|^2}.
  \label{extra1}
  \end{eqnarray}
The purple and green NLS orbits correspond, respectively, 
to typical oscillating and rotating orbits.
Aligned with the stability results, the reduced ODE model accurately
captures the nonlinear evolution for these orbits. This again supports
the conclusion that the reduced ODE is an accurate, qualitative and
quantitative, depiction, not only for the statics and stability (as
seen earlier), but also for the dynamics of vortex orbits in the torus.

In Fig.~\ref{fig:dip_dynamics} we depict the dynamics ensuing from the
destabilization of unstable dipole configurations. Specifically, the
top left and right panels depict one period for, respectively, an
oscillating and a rotating orbit.
As discussed before, the horizontal dipole-in steady state corresponds to 
a saddle (cf.~phase spaces of Fig.~\ref{fig:phase_space_hor_dip}) whose
separatrices separate regions with oscillating and rotating orbits.
The oscillating orbit was obtained by slightly and symmetrically perturbing
the vortices in the poloidal direction. Similarly, the rotating orbit
was obtained by slightly and symmetrically perturbing in the toroidal direction.
The bottom panel in Fig.~\ref{fig:dip_dynamics} depicts a typical
destabilization of the diagonal dipole. In this case, as the symmetry is
already broken from the steady state, the destabilization dynamics
follows windings along both toroidal and poloidal directions. Since the
diagonal dipole has an angle that is close to horizontal (toroidal), the
dipole has a relatively fast poloidal speed and a relatively slow toroidal drift.
The ensuing orbit will be generically a quasi-periodic orbit (unless the
windings along toroidal and poloidal are commensurate to each other).

\begin{figure}[t]
\includegraphics[width=1.00\columnwidth]{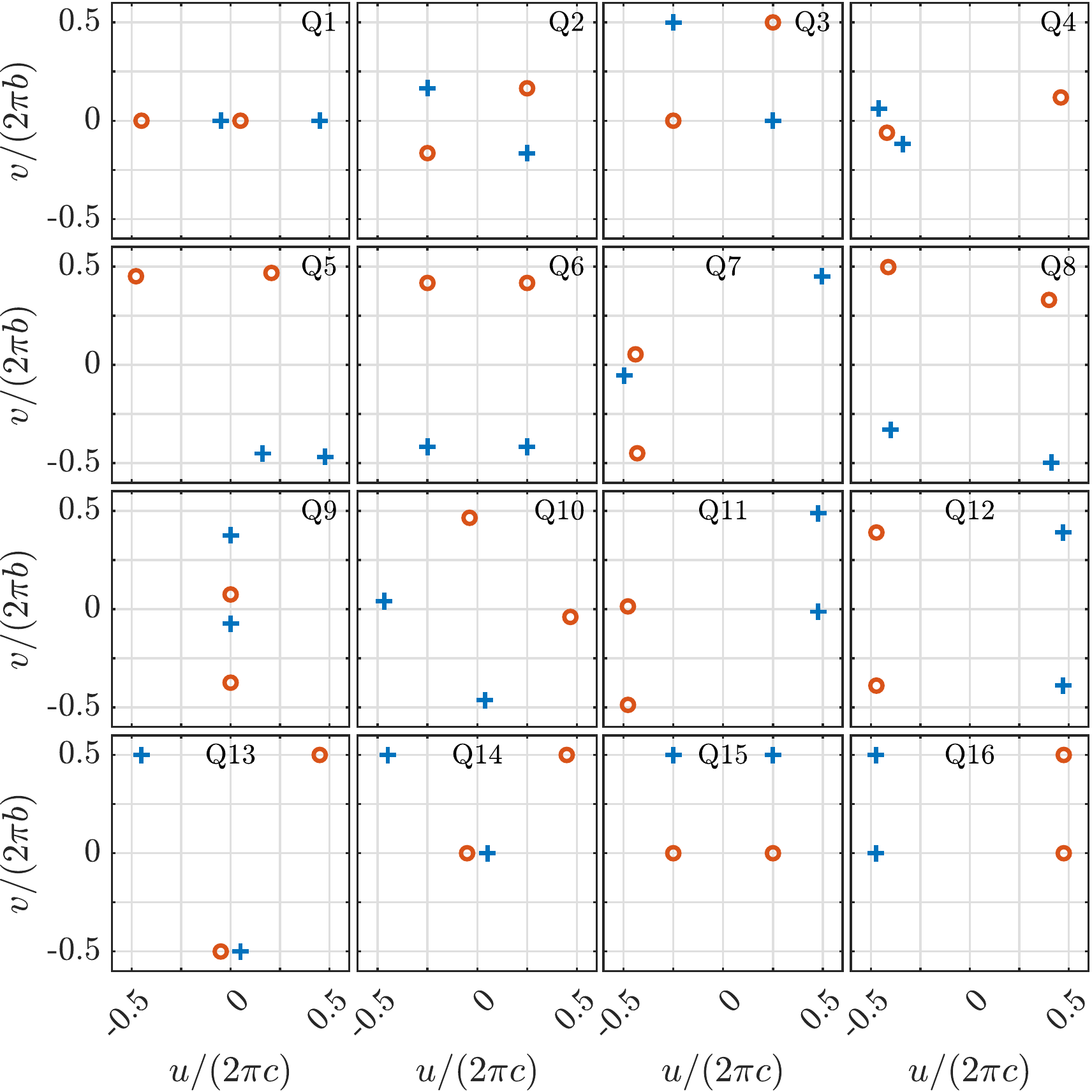}
\caption{(Color online)
Quadrupole solutions from the ODE model for $R=12$ and $\alpha=0.5$.
These solutions are ordered from least unstable to most unstable
(see Fig.~\ref{fig:quads_evals} for the corresponding spectra).
Positive (negative) vortices are depicted with the blue (red)
crosses (circles).
Vortex locations are plotted in scaled isothermal coordinates.
}
\label{fig:quads_configs}
\end{figure}

\subsection{Vortex Quadrupoles}
\label{sec:num:quad}

\subsubsection{Vortex Quadrupoles: steady states}
\label{sec:num:quad:steady}

While there exist only four steady state vortex dipoles, as the number
of vortices is increased, a larger zoo of possibilities arises.
Motivated by the remarkable agreement of the reduced ODE model with the 
original NLS, we have proceeded to leverage its use to identify possible quadrupole 
solutions in the full NLS model.
Even when using the reduced ODE model,
an exhaustive (ordered) search for quadrupoles (and higher order tuples),
as it was performed for the vortex dipoles-in Sec.~\ref{sec:num:dip}, 
is a challenging task.
This is because of the commonly referred to ``curse of dimensionality''.
While for the vortex dipole, after eliminating the toroidal translational 
invariance, the reduced ODE model is left with three degrees of freedom, for 
the vortex quadrupole one has (after eliminating the translational invariance)
seven degrees of freedom. Therefore, an exhaustive search over the
whole phase space is computationally prohibitive. Thus, we revert
to randomly sampling initial conditions over this seven-dimensional 
space (for all other parameters fixed; namely $R$ and $\alpha$)
and using a standard fixed point iteration (nonlinear least squares
with a Levenberg-Marquardt algorithm) to converge to the `closest'
steady state solution. 
Using several million initial conditions we were able to detect 16 distinct
quadrupole configurations as depicted in Fig.~\ref{fig:quads_configs}
for $R=12$ and $\alpha=0.5$.
By distinct we mean here that we have eliminated all the equivalent
solutions (not only through toroidal translations but also) through
symmetries associated with reflections across $\theta=0$, symmetries
associated with reversing the vortex charges, and permutations of the 
vortex labels.
We have therefore obtained a rich palette of quadrupole solutions as depicted 
in Fig.~\ref{fig:quads_configs}. It is worth mentioning that these solutions 
have been ordered Q1, ..., Q16 from the least unstable to the most unstable 
one (for the parameters at hand; namely $R=12$ and $\alpha=0.5$).
They include horizontal (Q1) and vertical aligned quadrupoles (Q9),
rectangular quadrupoles (Q2, Q3, Q6, Q12, and Q16),
rhomboidal quadrupoles (Q8, Q10, Q11, and Q15),
trapezoidal quadrupoles (Q5, Q13, and Q14),
and, somewhat surprisingly, irregular quadrupoles (Q4 and Q7).

\begin{figure}[t]
\includegraphics[width=1.00\columnwidth]{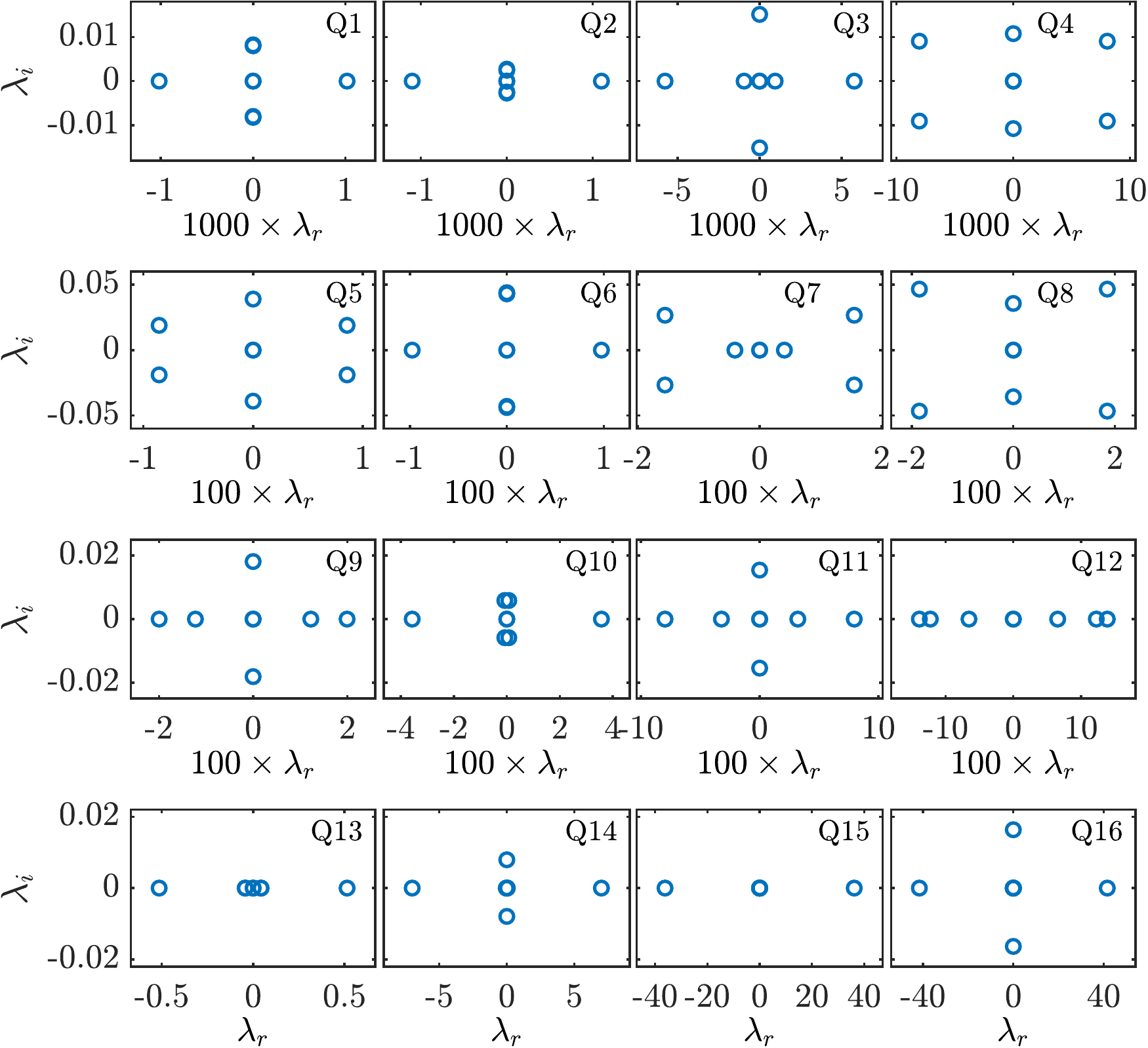}
\caption{(Color online)
Stability spectra $(\lambda_r,\lambda_i)$ corresponding to the 
quadrupole solutions of Fig.~\ref{fig:quads_configs} obtained
from the reduced ODE model.
Note that each row of panels has a different scaling for
the real part of the eigenvalue as indicated.
The configurations are ordered from least unstable to most unstable.
In particular, the first three quadrupoles present a very weak
instability of $O(10^{-3})$ for this value of the parameters
($R=12$ and $\alpha=0.5$).
}
\label{fig:quads_evals}
\end{figure}

\begin{figure}[htb]
\includegraphics[width=0.90\columnwidth,height=3.6cm]{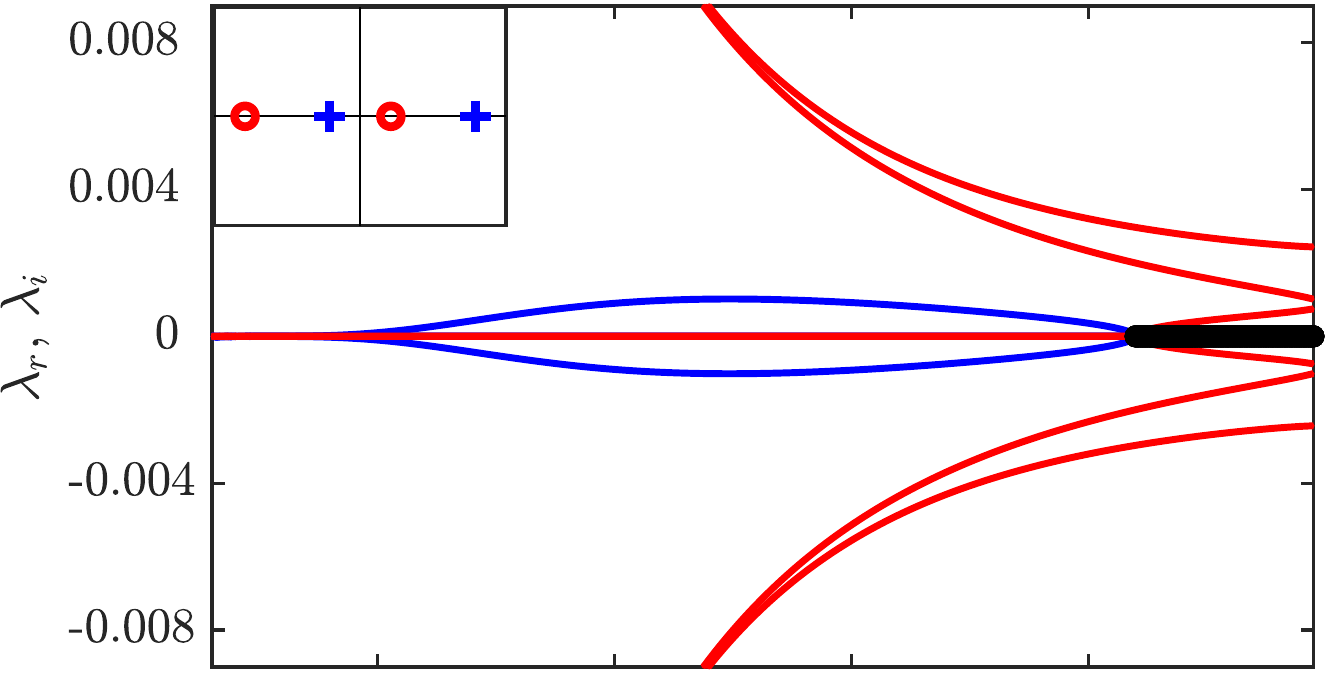}
\\[1.0ex]
\includegraphics[width=0.90\columnwidth,height=3.6cm]{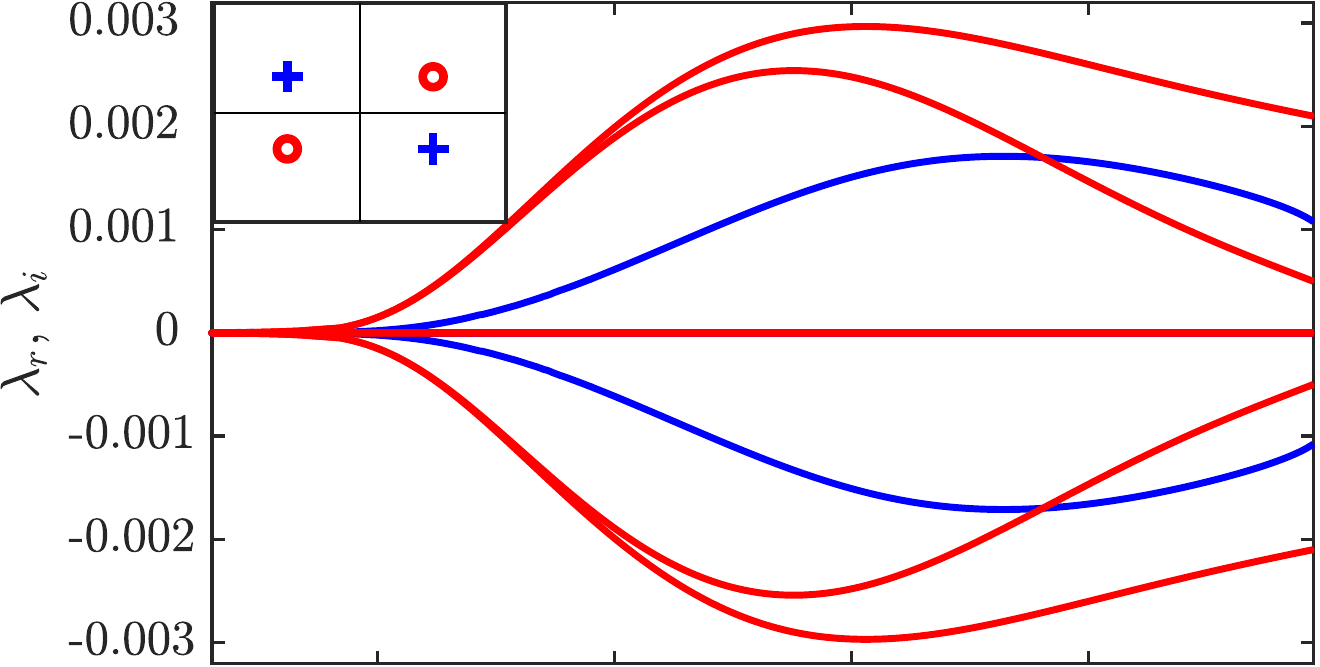}
\\[0.7ex]
\includegraphics[width=0.90\columnwidth,height=4.3cm]{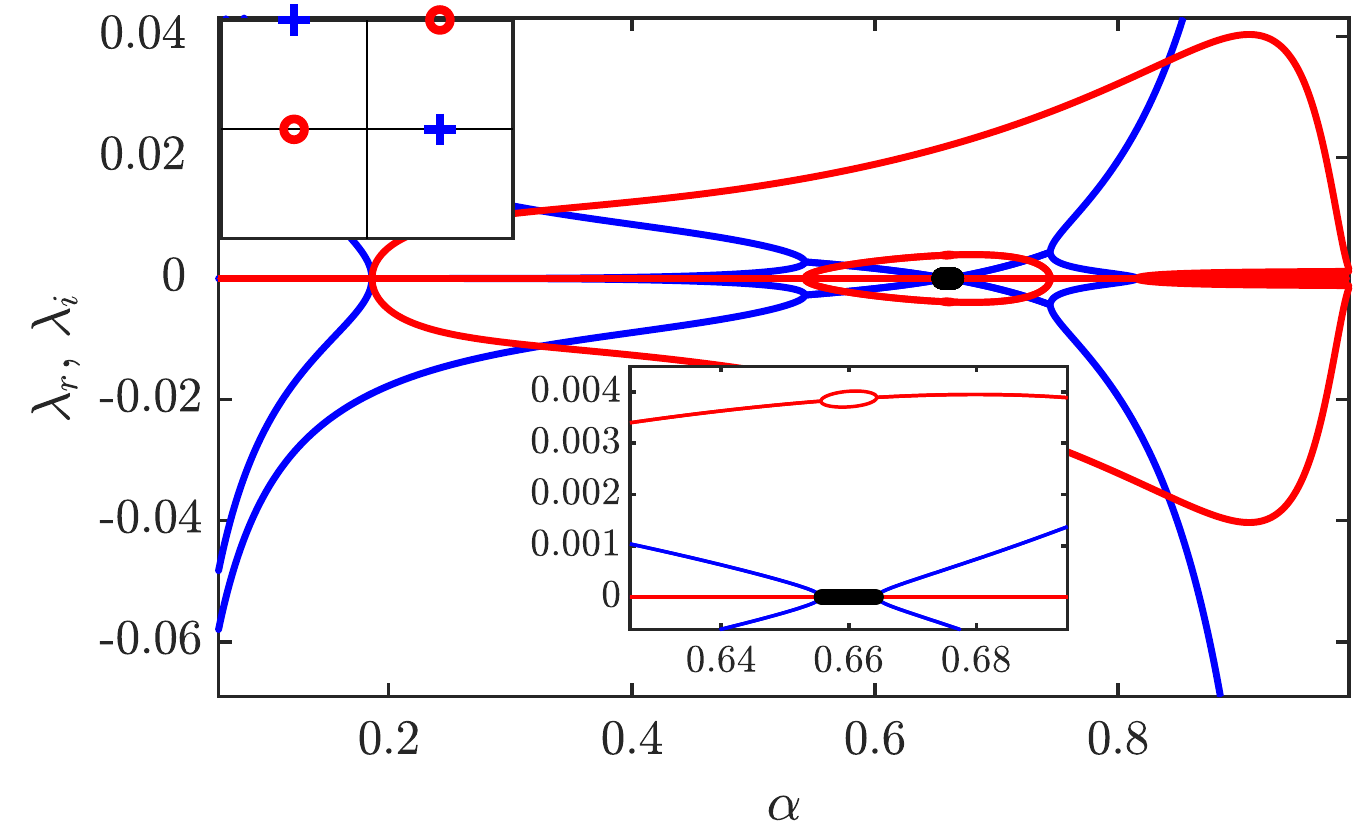}
\caption{(Color online)
Stability spectra as $\alpha$ is varied in the reduced ODE model for the 
first (Q1; top), second (Q2; middle) and third (Q3; bottom) quadrupole solutions 
(see Fig.~\ref{fig:quads_configs}) for $R=12$.
Blue and red curves correspond to the real ($\lambda_r$) and 
imaginary ($\lambda_i$) parts of the eigenvalue.
The top-left insets depict the configurations for $\alpha=0.5$.
The Q1 and Q3 quadrupole solutions were the only ones found to have
stability intervals as $\alpha$ was varied. These stability
intervals are depicted by the thick black lines.
The bottom inset for Q3 corresponds to a zoom around the
stability interval.
}
\label{fig:stability_Q1_Q2_Q3}
\end{figure}

\begin{figure}[htb]
\includegraphics[width=0.99\columnwidth]{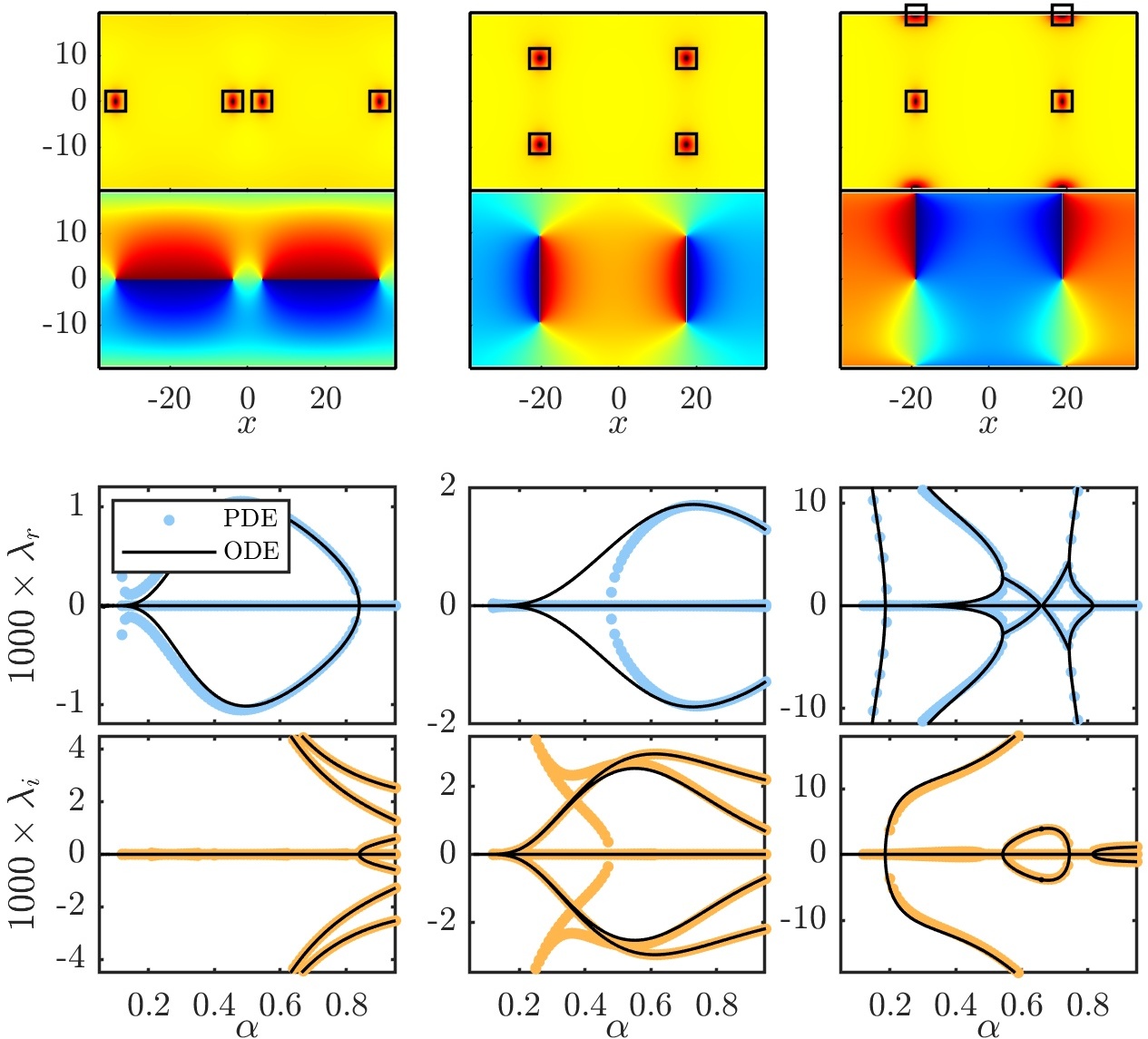}
\caption{(Color online)
Comparing full model (PDE) and reduced model (ODE) spectra for quadrupoles 
Q1 (left), Q2 (middle), and Q3 (right) as $\alpha$ is varied for $R=12$ 
and $\mu=1$.
Similar layout as in Fig.~\ref{fig:spectra_alpha_mu1}.
}
\label{fig:stability_PDE_Q1_Q2_Q3}
\end{figure}

\begin{figure}[htb]
\includegraphics[width=0.95\columnwidth]{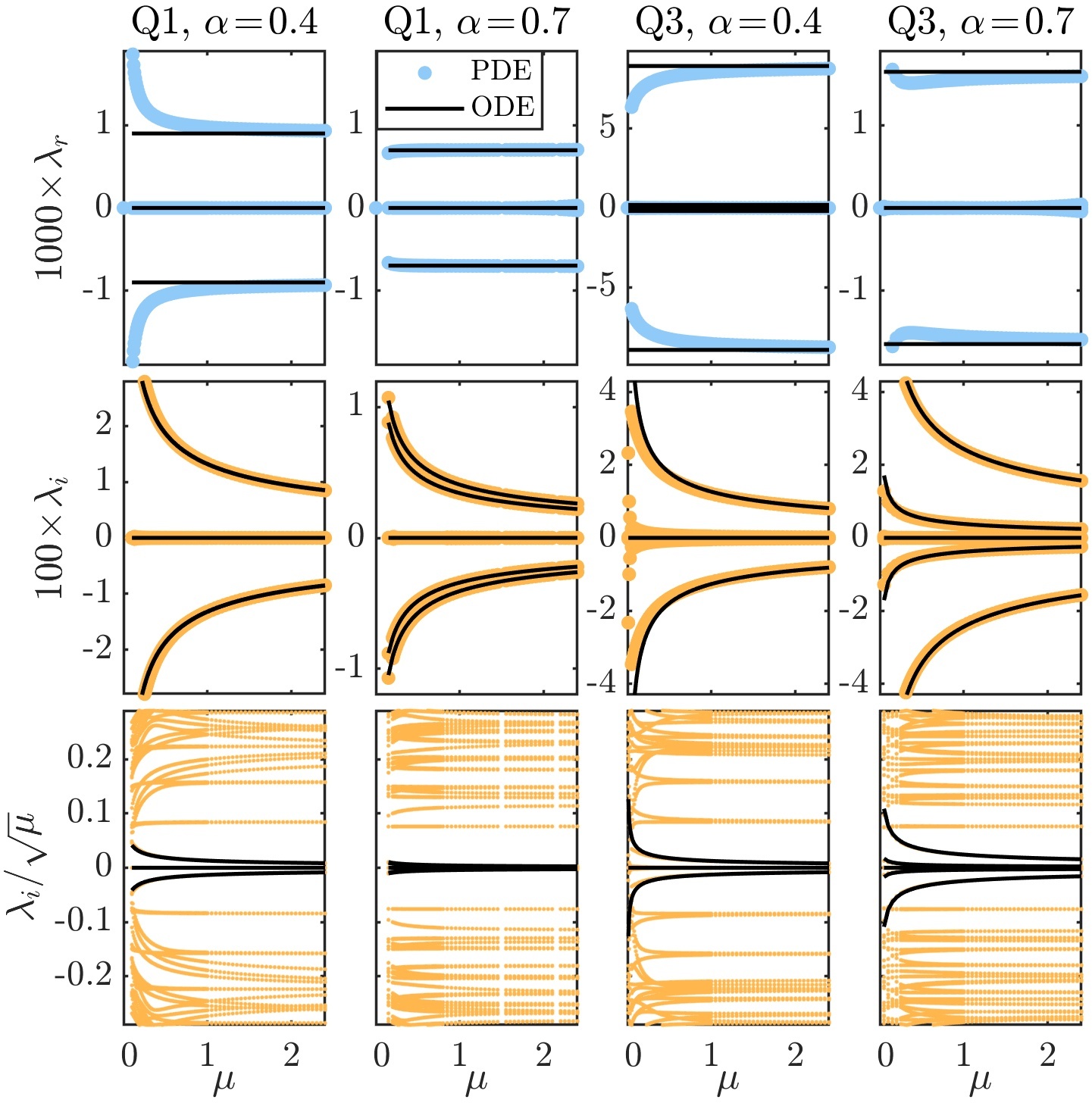}
\caption{(Color online)
Convergence of the stability spectra for the Q1 and Q3
quadrupole configurations as the chemical potential $\mu$ is increased 
for $R=12$ and $\alpha=0.4$ and $\alpha=0.7$ as indicated.
Same layout as in Fig.~\ref{fig:spectra_mu_alpha04}.
}
\label{fig:spectra_Q1Q3_vs_mu}
\end{figure}

\subsubsection{Vortex Quadrupoles: stability}
\label{sec:num:quad:stab}

Let us now comment on the stability for the obtained quadrupole solutions.
We  start by analyzing the stability obtained from the reduced ODE model.
Figure~\ref{fig:quads_evals} depicts the ODE spectra associated with the
16 distinct quadrupoles depicted in Fig.~\ref{fig:quads_configs}.
As mentioned above, the solutions have been ordered from least unstable to most 
unstable by using the maximal real part for all eigenvalues $\max(\lambda_r)$.
It is important to stress that the eigenvalues, and thus the ordering of the 
quadrupole solutions as we have posited it, changes as the parameters of the system
are varied. 
Therefore, it is relevant to look at the associated spectra as the
parameters are varied. Particularly revealing for some of these solutions 
are the spectra continuation as the torus aspect ratio $\alpha$ is varied.
For compactness, we only show these full ODE spectra for the first 
three quadrupole configurations in Fig.~\ref{fig:stability_Q1_Q2_Q3}. 
These configurations, as shown in Fig.~\ref{fig:quads_evals}, have a very weak 
instability ($\max(\lambda_r)<0.01$) for $\alpha=0.5$ (and $R=12$) and thus 
are potential candidates to be completely stable as the parameters are varied. 
As Fig.~\ref{fig:stability_Q1_Q2_Q3} shows, for $R=12$, Q2 is always unstable, 
however this instability is rather weak as it merely reaches $\max(\lambda_r)\simeq 0.0017$
around $\alpha\simeq0.73$ 
On the other hand, Q1 and Q3 are not only weakly unstable for most values of
$\alpha$, but, importantly, they can be rendered {\em stable} on respective
windows of the parameter $\alpha$. 
Specifically, as the figure shows, for $R=12$, the Q1 solution is stable for
$0.84\lesssim \alpha <1$ and the Q2 solution is stable for the narrow 
interval $0.656\lesssim\alpha\lesssim 0.664$ (see bottom inset for Q3).
In addition to the stability windows for Q1 and Q3, it is also worth mentioning that 
the reduced model predicts that a few quadrupole solutions have relatively weak
instabilities. For instance, as it can be seen in Fig.~\ref{fig:quads_evals}, the quadrupoles Q1 and Q2 have a $\max(\lambda_r)\simeq 10^{-3}$,
while Q3 has a $\max(\lambda_r)\simeq 6\times 10^{-3}$, and
while Q4 has a $\max(\lambda_r)\simeq 9\times 10^{-3}$, all for
$\alpha=0.5$ and $R=12$.
Therefore, for the parameter combinations that we explored, although we only found 
Q1 and Q3 to possess stability windows, the rather weak instabilities presented by 
about half of the quadrupole configurations ($\max(\lambda_r)<0.02$), suggest that 
they have the potential to be long-lived solutions in the original NLS equation
(cf.~dynamics for Q1 in Sec.~\ref{sec:num:quad:dyn})
and the corresponding physical experiments.

We now turn to the study of the stability of the quadrupole configurations from
the full NLS model and compare the latter  with the
results from the reduced ODE model.
In particular, Fig.~\ref{fig:stability_PDE_Q1_Q2_Q3} shows  the stability results 
from  both the reduced ODE and the full NLS models for the Q1, Q2, and Q3
configurations as $\alpha$ is varied for $R=12$ and $\mu=1$. 
It is remarkable that, even with this moderately small value of $\mu$ ($\mu=1$),
the reduced ODE model is capable to recover the main stability eigenvalues of
the full NLS for most parameter values.
For instance, the complicated bifurcation scenario displayed by Q3 is
perfectly captured by the ODE model (including the bifurcation leading
to the narrow stability window around $\alpha \simeq 0.66$).
In general, we expect that the ODE model is able to properly capture the 
instabilities corresponding to the destabilization of the vortex positions.
However, it is conceivable that there exists instabilities for full NLS 
solutions that are not captured by the reduced ODE model as the corresponding 
eigendirections might not be part of the space spanned by the latter. 
Surprisingly however, as it can be observed from the spectrum of Q2 in 
Fig.~\ref{fig:stability_PDE_Q1_Q2_Q3}, the NLS configurations seems to 
{\em stabilize} for $\alpha\lesssim 0.48$ while the reduced ODE model 
predicts that the solution is (weakly) unstable for all values of $\alpha$.
This tends to suggest that the actual spatial extent of the vortices, and 
their mutual interaction through the curvature of the background, may be
responsible for this stabilization since the point-vortex model, relevant for 
the $\mu\rightarrow\infty$ limit, does not display this stabilization effect.

Finally in Fig.~\ref{fig:spectra_Q1Q3_vs_mu} we depict the convergence of
the stability spectra for the Q1 and Q3 dipoles as $\mu$ is increased for
$R=12$ and $\alpha=0.4$ and $\alpha=0.7$.
As it was the case for the vortex dipoles, the convergence between the
original NLS and the reduced ODE model as $\mu$ increases is extremely good.
In fact, even for moderate values of $\mu\approx 2$ the discrepancy between 
the maximum real parts of the eigenvalues for the full NLS and the ODE model
is less than $4\%$ in all the examined cases. Furthermore, the
qualitative stability conclusions do not appear to change over $\mu$
for the cases and intervals considered.

\begin{figure}[t]
\includegraphics[width=4.1cm,height=2.5cm]{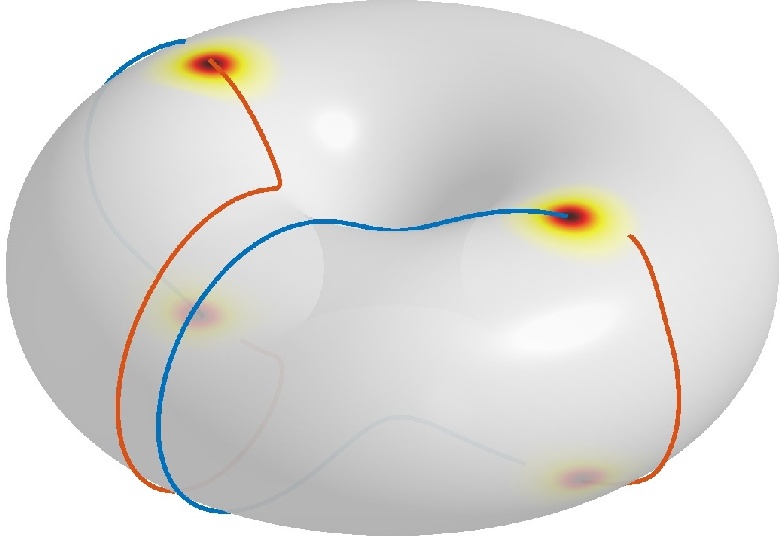}
~
\includegraphics[width=4.1cm,height=2.5cm]{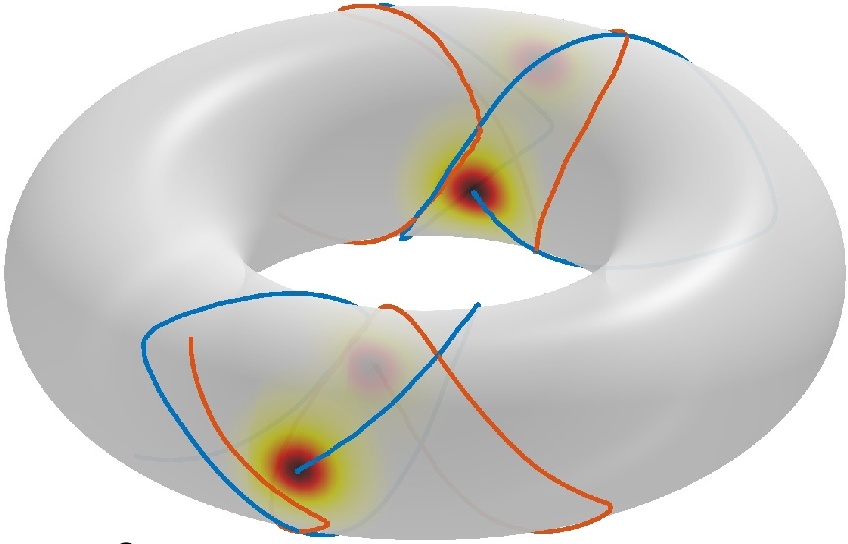}
\\[2.0ex]
\includegraphics[width=4.1cm,height=2.5cm]{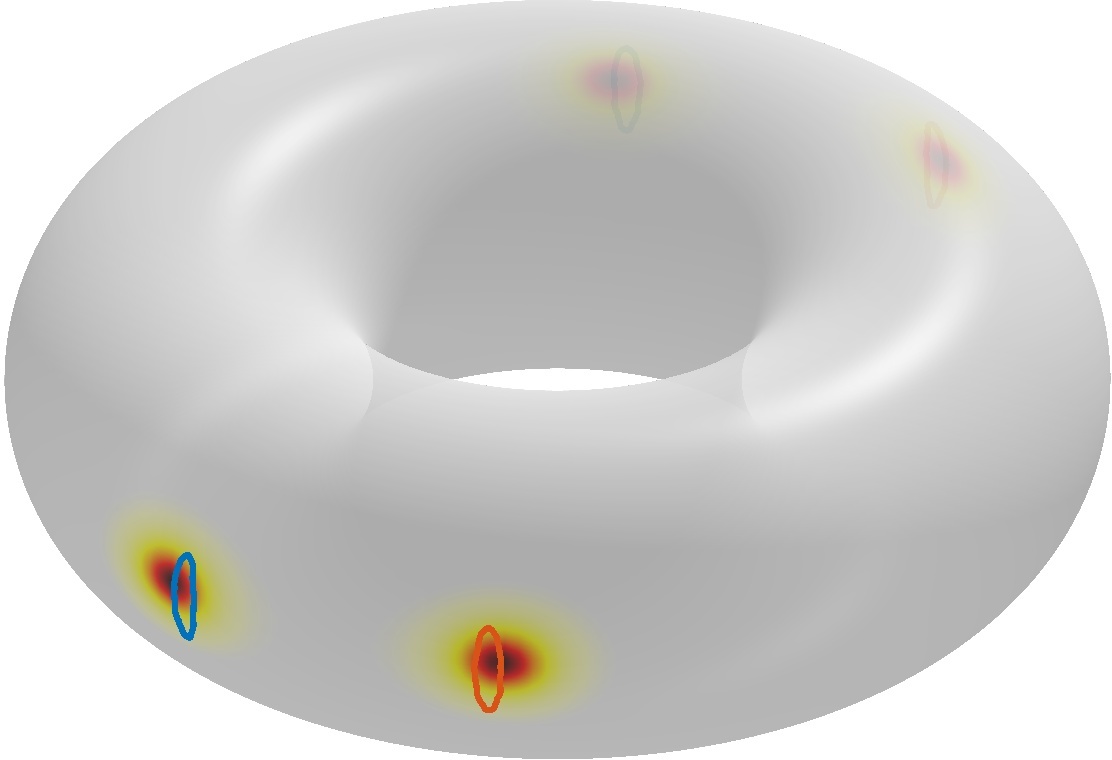}
~
\includegraphics[width=4.1cm,height=2.5cm]{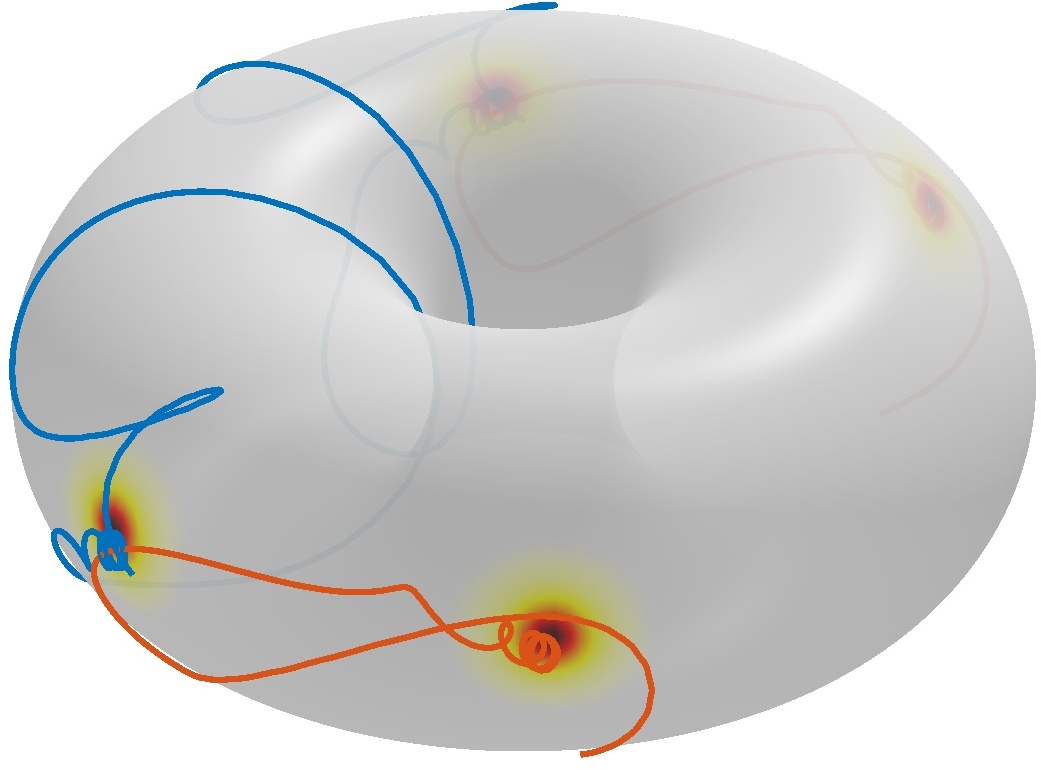}
\caption{(Color online)
Dynamics ensuing from the destabilization of unstable quadrupole configurations
for $R=12$ and $\mu=1$.
Top-left: destabilization of the Q2 quadrupole for $\alpha=0.7$ giving rise 
to approximately (about half of) a periodic orbit.
Top-right: destabilization of the Q3 quadrupole for $\alpha=0.4$.
Bottom-left: apparently stable periodic orbit for a perturbed Q1 quadrupole 
for $\alpha=0.5$ for $0\leq t \leq 4000$.
Bottom-right: the apparently stable periodic orbit for a perturbed Q1 quadrupole 
for $\alpha=0.7$ eventually destabilizes for longer times $0\leq t \leq 28000$.
The colored surface depicts the initial density and the overlaid curves
correspond to the trajectory traces from the negative (red) and positive 
(blue) vortices.
Please see 
{\tt Q2-R12-alpha07-mu1.gif}, 
{\tt Q3-R12-alpha04-mu1.gif}, 
{\tt Q1-R12-alpha05-mu1.gif}, and
{\tt Q1-R12-alpha07-mu1-t28K.gif}
in the Supplemental Material for respective
movies depicting the evolution of the density and phase.
}
\label{fig:quad_dynamics}
\end{figure}

\subsubsection{Vortex Quadrupoles: dynamics}
\label{sec:num:quad:dyn}

Let us now follow the destabilization dynamics for quadrupole configurations. 
Figure~\ref{fig:quad_dynamics} depicts the dynamics ensuing for the 
first three quadrupole steady state configurations (Q1, Q2, and Q3).
In particular, the top-left panel depicts the destabilization of the Q2
dipole. The corresponding orbit follows the unstable manifold of this
unstable saddle fixed point and seems to return along the stable manifold
suggesting, as it was the case for the horizontal dipoles 
(cf.~Fig.\ref{fig:phase_space_hor_dip}), the existence of a homoclinic orbit.
The top-right panel depicts a typical destabilization for the Q3 dipole.
Finally, the bottom panels depicts the dynamics ensuing from perturbing
the Q1 quadrupole. It is important to stress that the values of $\alpha$
($\alpha=0.5$ for the left panel and $\alpha=0.7$ for the right panel)
are {\em below} the stability window that starts around $0.84\lesssim \alpha$.
Therefore, for these values of the parameters, the Q1 quadrupole is unstable.
However, when following the perturbed dynamics for times of the order of 
several thousands (see bottom-left panel) the vortices seem to trace a 
neutrally stable periodic (center) periodic orbit. The reason for this 
apparent stability stems from the fact that the Q1 quadrupole is unstable but 
that its instability eigenvalue is rather weak ($\max(\lambda_r)\simeq0.001$; 
see top panel in Fig.~\ref{fig:stability_Q1_Q2_Q3}).
The instability for the Q1 quadrupole is indeed manifested for longer
times as shown in the bottom-right panel where the vortices slowly begin
to spiral out for $t>10,000$ and finally engage in an apparent irregular
dance for $t>15,000$.

Although in the case of the dipole the motions we have encountered seem
to be prototypically quasi-periodic, in the case of quadrupole configurations,
there are six degrees of freedom and despite the presence of a conserved energy,
there is potential for complex (chaotic) behavior. 
Studies along these lines are outside the scope of the present paper
and will be presented in future work.

\begin{figure}
\includegraphics[width=0.9\columnwidth]{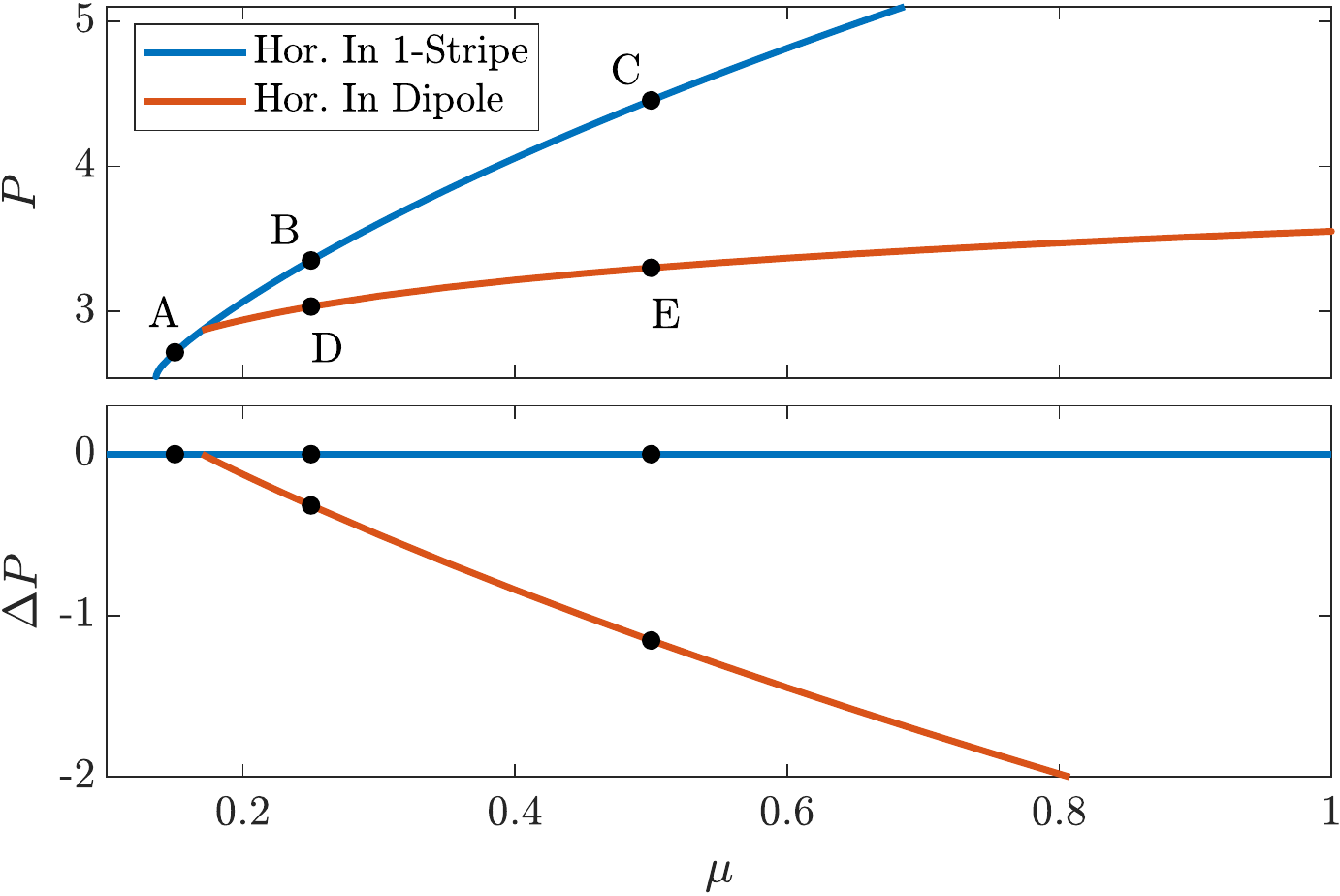}
\\[2.0ex]
\includegraphics[width=1.0\columnwidth]{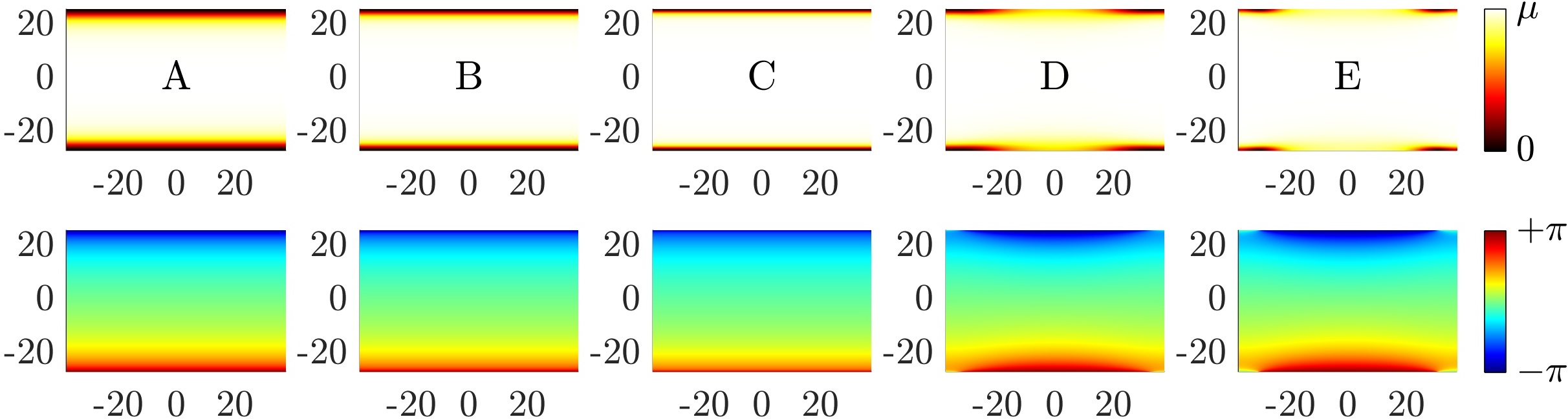}

\caption{(Color online)
Bifurcation between the horizontal-in single dark soliton stripe and 
the horizontal-in dipole.
The top panel depicts the power for both solutions as a function
of $\mu$ for $\alpha=0.7$ and $R=12$.
The second panel depicts the power difference between the two solutions.
The last two row of panels depict the density (top subpanels) and
phase (bottom subpanels) of the solutions in Cartesian coordinates
at the points indicated in the first panel. Namely:
$\mu=0.15$ for point A,
$\mu=0.25$ for points B and D, and
$\mu=0.5$ for points C and E.
}
\label{fig:bif_HI_stripe}
\end{figure}

\begin{figure}[htb]
\includegraphics[width=0.85\columnwidth]{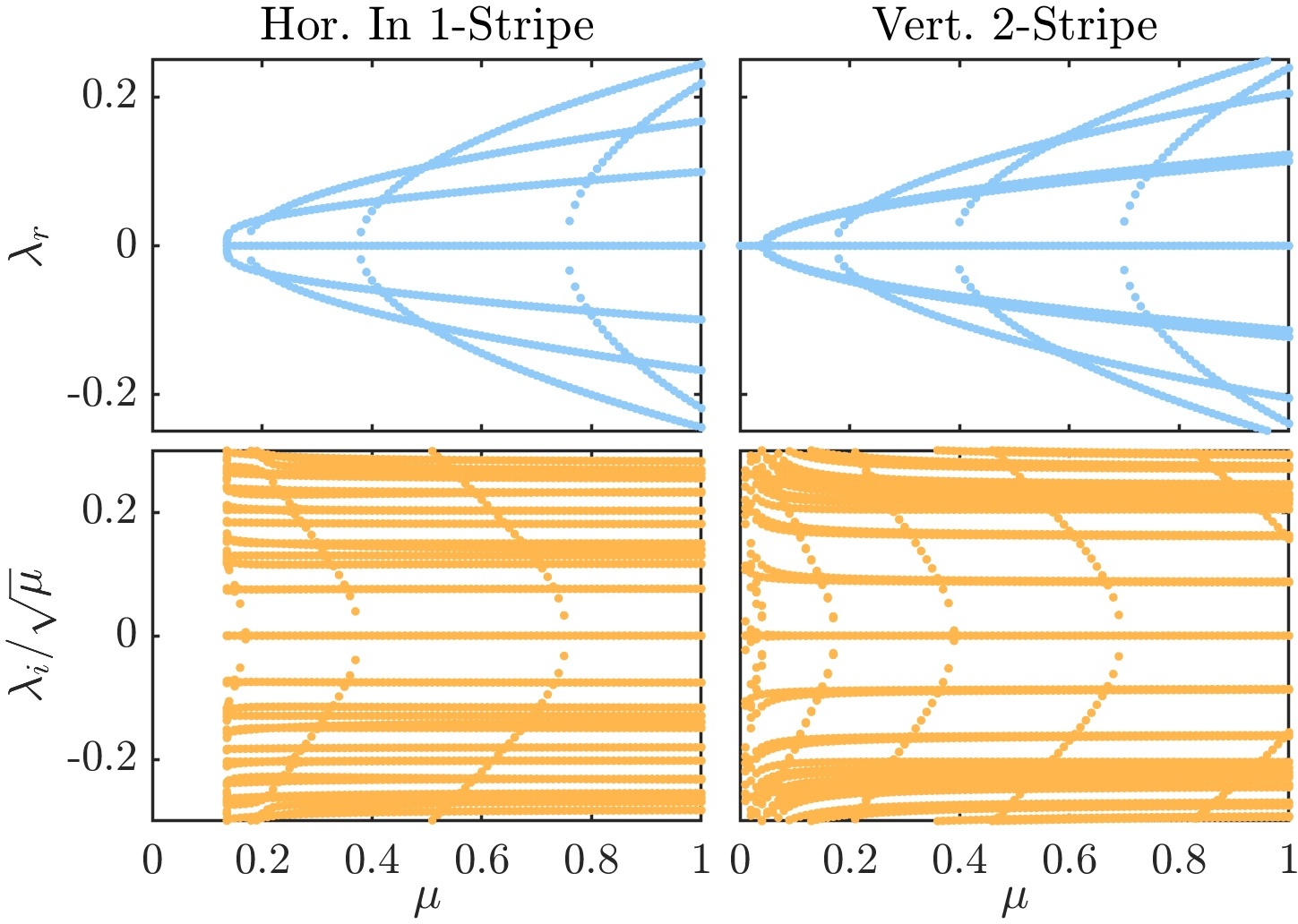}
\caption{(Color online)
Bifurcation spectra for the horizontal-in single stripe for $\alpha=0.7$ (left) 
and the vertical double stripe for $\alpha=0.4$ (right).
Sample instability eigenfunctions are shown in Fig.~\ref{fig:evec_2SVmu0p1}.
}
\label{fig:evals_HI1S_V2S}
\end{figure}

\subsection{Dark Soliton Stripes}
\label{sec:num:DSSs}

Finally, let us briefly examine some of the properties of dark soliton
stripe configurations and their connection with some of the vortex 
configurations of the previous sections.
For instance, as Fig.~\ref{fig:HIsamples} suggest, as $\mu$ is decreased, the 
horizontal dipole-in seems to degenerate, as the vortices eventually merge, 
into a {\em single} horizontal (toroidal) stripe. 
It is important to first understand the nature of this single stripe.
As is the case for vortices, in order to satisfy the periodicity of
the domain, dark soliton stripes should appear in pairs as each one
will contribute to a $\pi$ phase jump for a total of a $2\pi$ phase
jump conducive to a periodic domain. However, it is possible to seed
a {\em single} dark soliton if one adds a winding to add (or counter)
the $\pi$ phase jump of a single stripe. Thus, it is in principle possible
to construct single horizontal stripe configurations that 
are consonant the periodicity by adding a vertical winding 
$W_y=n+1/2$ with $n$ integer.
Likewise single vertical  stripes are possible when adding a half-integer
horizontal winding $W_x=n+1/2$. In fact, it is possible to construct any
number of {\em odd} stripes by adding half-integer windings. On the
other hand, similarly to our discussion for vortices, from the
perspective of the torus periodicity, it is possible
for our system to feature even numbers of parallel dark stripes (without
external winding) whose total phase change adds up to a multiple of $2 \pi$.

In Fig.~\ref{fig:bif_HI_stripe} we depict more details of the
bifurcation between the horizontal-in single stripe and the corresponding dipole solutions.
In particular, we follow the effective power of the solutions
by computing:
\begin{eqnarray}
  P = \iint_0^{2\pi}\left(\mu- |\psi|^2\right)\, dS,
  \label{norm}
\end{eqnarray}
where the surface element on the torus is 
$dS= |\hat{\phi}\times\hat{\theta}| d\theta d\phi$.
Here hats denote the unit vectors in the different (toroidal and poloidal) 
directions. Thus, $|\hat{\phi}\times\hat{\theta}| = R + r\cos(\theta)$.
As the top two panels of Fig.~\ref{fig:bif_HI_stripe} evidenced,
the horizontal-in dipole bifurcates from the horizontal-in single
stripe at $\mu\approx 0.17$.
The bottom two rows of panels in Fig.~\ref{fig:bif_HI_stripe} depict sample 
solutions before and after the bifurcation. The phase panels indicate that 
the stripe has indeed an extra vertical winding of $W_y=-1/2$ and
that the vortices of the dipole merge as $\mu$ decreases towards
the bifurcation at $\mu\approx 0.17$.

\begin{figure}[t]
\includegraphics[width=1.0\columnwidth]{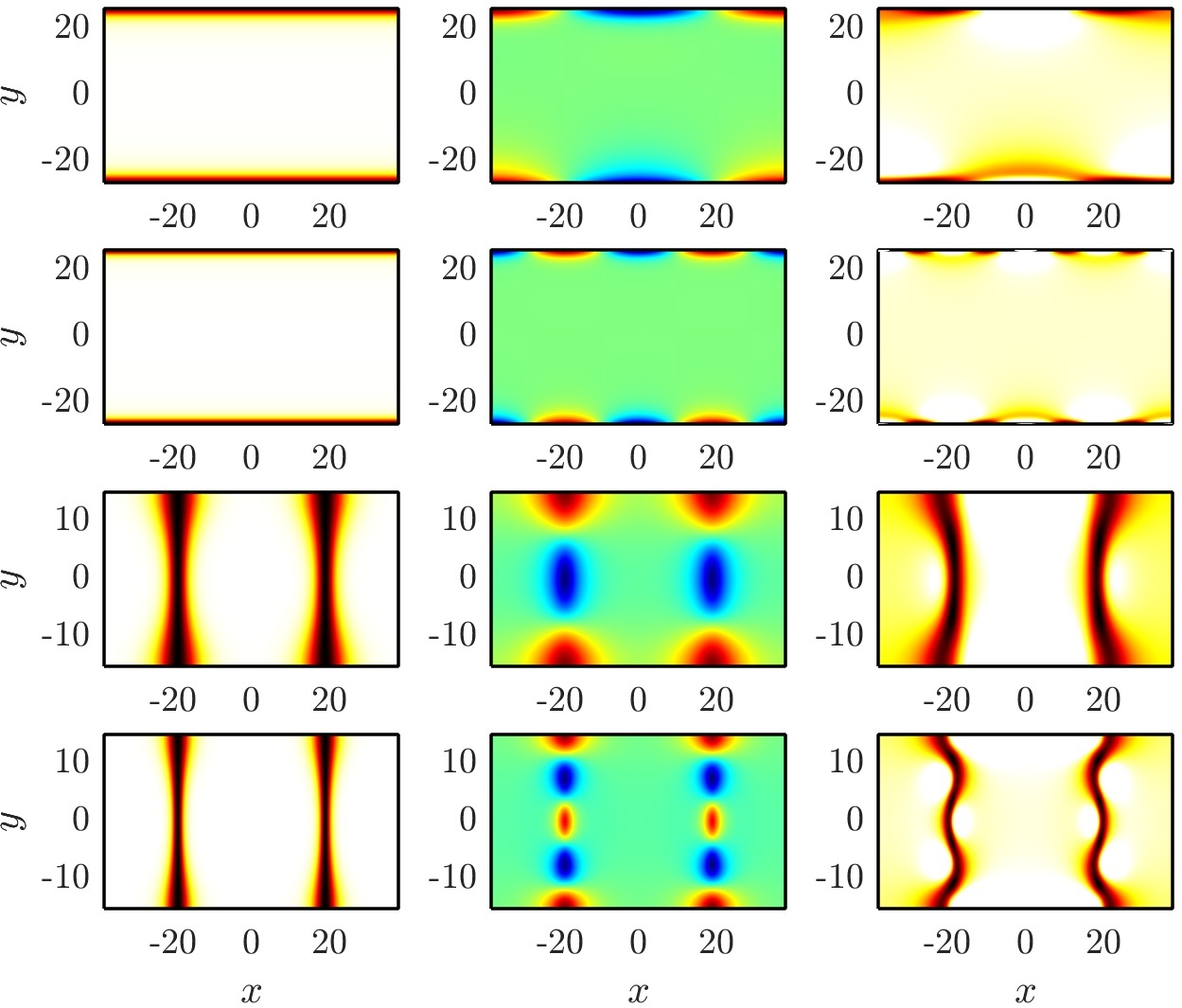}
\caption{(Color online)
Effects of the most unstable eigenfunctions for stripe configurations for $R=12$.
The left panels depict the steady state configuration while the middle
panels their corresponding (real part of the) most unstable eigenfunction.
The right panels depict the stripe configuration after a (large)
perturbation with the most unstable eigenfunction.
All states are plotted in Cartesian coordinates.
First and second rows: 
horizontal-in single stripe for $\alpha=0.7$ and $\mu=0.25$ and $\mu=0.5$, respectively. 
Third and fourth rows:
vertical two-stripe for $\alpha=0.4$ and $\mu=0.1$ and $\mu=0.3$, respectively. 
The corresponding evolution dynamics (for a much smaller perturbation
strength) are depicted in Fig.~\ref{fig:run_stripes}.
}
\label{fig:evec_2SVmu0p1}
\end{figure}

\begin{figure*}[htb]
\includegraphics[width=2.04\columnwidth]{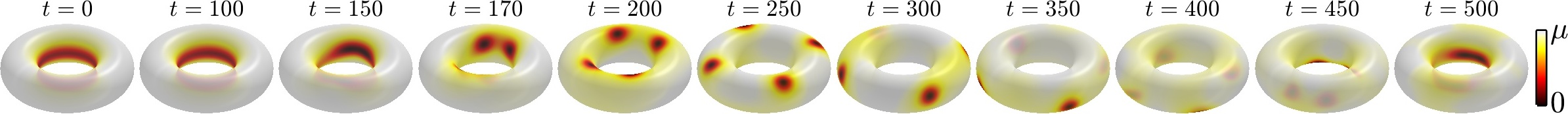}
\\[1.0ex]
\includegraphics[width=2.04\columnwidth]{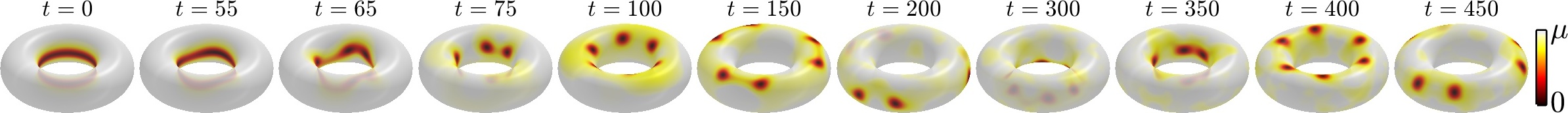}
\\[1.0ex]
\includegraphics[width=2.04\columnwidth]{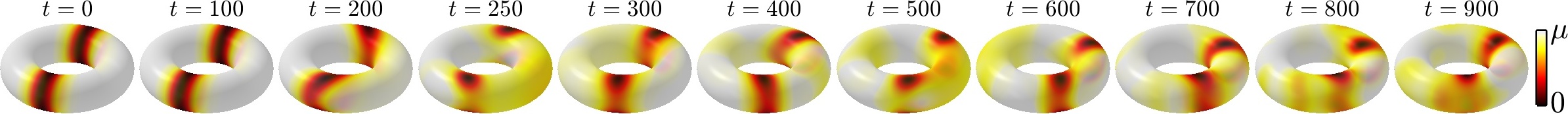}
\\[1.0ex]
\includegraphics[width=2.04\columnwidth]{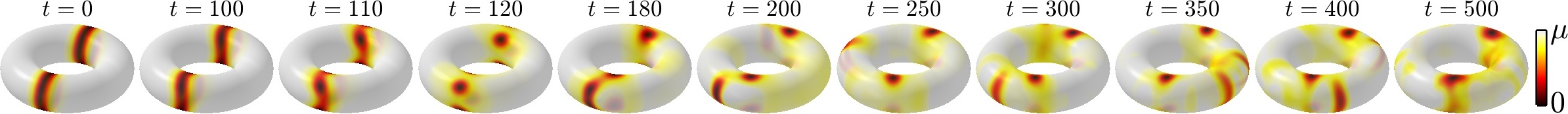}
\caption{(Color online)
Evolution dynamics of unstable stripe configurations for $R=12$.
All initial conditions are taken as the stationary stripe configuration
perturbed by $10^{-3}$ times the most instable eigenfunction.
First and second rows: 
horizontal-in single stripe for $\alpha=0.7$ and $\mu=0.25$ and $\mu=0.5$, respectively. 
Third and fourth rows:
vertical two-stripe for $\alpha=0.4$ and $\mu=0.1$ and $\mu=0.3$, respectively. 
Please see 
{\tt 1SHImu0p25.gif}, 
{\tt 1SHImu0p50.gif}, 
{\tt 2SVer-R12-alpha04-mu0.1.gif}, and
{\tt 2SVer-R12-alpha04-mu0.3.gif}, 
in the Supplemental Material for respective
movies depicting the evolution of the density and phase.
}
\label{fig:run_stripes}
\end{figure*}

In a similar fashion as the horizontal dipole-in bifurcates from the 
single horizontal-in stripe, multiple other bifurcations are present 
involving single and double (and triple, etc.), in and out, vertical 
and horizontal dark stripes and vortex configurations. 
In fact, we have observed (not shown here) mixed bifurcations where,
as $\mu$ increases, a double horizontal stripe, containing an in and an out 
stripe, first features a bifurcation towards a mixed state containing a 
vortex out configuration and a stripe in which, subsequently, after further increase 
in $\mu$, displays a bifurcation for the stripe in leading to a
multi-vortex state.
An in-depth analysis of the possible bifurcations involving the above
configurations, albeit interesting, falls 
outside of scope of the present work and will be reported elsewhere.
Nonetheless, we present here a collection of examples that showcase
some of the most basic bifurcations and ensuing dynamics involving dark
soliton stripes.

Figure~\ref{fig:evals_HI1S_V2S} depicts the bifurcation spectra
for the horizontal-in single stripe (left) and the vertical double stripe (right).
The spectra for both stripe configurations display a series (cascade) of
bifurcations as $\mu$ departs from zero. Each bifurcation is associated with
the creation of an offspring configuration where each stripe is replaced by 
vortices. The higher the bifurcation in the cascade the more vortices are produced.
This cascading bifurcation is akin to the bifurcation of dark soliton
stripes and rings in parabolically trapped BECs as reported, e.g., in 
Ref.~\cite{Middelkamp:2011}.
In Fig.~\ref{fig:evec_2SVmu0p1} we portray elements of these bifurcations by
perturbing the stripe steady states by the eigenfunction corresponding to the 
most unstable eigenvalue as shown in Fig.~\ref{fig:evec_2SVmu0p1}.
The figure depicts the stripe steady states (left), their corresponding
most unstable eigenfunction (middle), and the steady state perturbed by the 
eigenfunction (right). In these cases we normalized the eigenfunction such that 
its maximum density (norm) coincided with the chemical potential $\mu$ of the 
steady state and we added it using a large perturbation prefactor equal to two.
This was done for presentation purposes to exaggerate the visual effects
of the perturbation (for the actual dynamical runs presented in 
Fig.~\ref{fig:run_stripes} we used a small prefactor for the relevant
perturbation of $10^{-3}$).
As Fig.~\ref{fig:evec_2SVmu0p1} shows, the different eigenfunctions
bend the stripes into snaking modes with a higher number of relevant undulations
as $\mu$ increases and the higher bifurcations in the cascade are reached.
Specifically, the first two rows of panels in the figure present the first 
two unstable modes of the horizontal-in single stripe each giving rise to an 
aligned quadrupole and hexapole, respectively, after the system is left to 
evolve as depicted in the first two rows of Fig.~\ref{fig:run_stripes}.
On the other hand, the third and fourth rows of panels in 
Fig.~\ref{fig:evec_2SVmu0p1} depict the first two excited modes of
the double vertical stripe which in this case give rise,
respectively, to a Q2-like quadrupole (see $t=250$) and an octupole
(see $t=120$). The corresponding
destabilization dynamics are shown in the third and fourth rows of
Fig.~\ref{fig:run_stripes}. 
Note that the destabilization along the second mode of the vertical double
stripe (fourth row of panels) initially generates a vortex octupole ($t=120$). 
However, as time progresses, the outer two vortex pairs on each side of the 
torus merge and produce two ``lumps'' that travel, in opposite directions, 
along the toroidal direction 
until they are eventually destroyed and contribute to background radiation.
These lumps correspond to solitonic structures dubbed Jones-Roberts 
solitons~\cite{Jones-Robert:1982} which have been observed in recent BEC 
experiments~\cite{bongs}. These are quite interesting to explore in their
own right in the realm of traveling solutions in the torus setting. While our
exploration herein has been restricted (due to their extensive wealth,
as we have tried to argue) to stationary states, it does not escape
us that such traveling waveforms, including ones involving vorticity
are of particular interest in their own right for future studies.

\section{Conclusions and Outlook}
\label{sec:conclu}

In this work we have attempted to give a systematic and extensive
(although by no means exhaustive) study of
the existence, stability and dynamics of dark  and vortical
structures in the nonlinear Schr\"odinger (NLS) equation on the surface of a 
torus as the torus aspect ratio $\alpha$ and the 
chemical potential $\mu$ of the solutions are varied.
We chiefly study vortex dipoles and quadrupoles but also touch upon dark soliton 
stripes and their connections to the former (through bifurcation cascades).
To gain insight into the statics, stability, and dynamics of vortex 
configurations on the full NLS model, we have importantly leveraged
the key insights offered by a remarkably successful (as we illustrate)
reduced particle model, introduced
in Ref.~\cite{fetter:torus}, based upon assuming vortices without internal
structure (i.e., point-vortices) that incorporates both vortex-vortex 
interactions and the effects of space curvature on the surface of the torus.
We also extend this reduced particle model to incorporate vertical
and horizontal phase windings that induce monotonic flows along the,
respectively, poloidal and toroidal directions of the torus.
We show that this (fundamental within this setting)
reduced particle model is extremely accurate at predicting
the statics, stability, and dynamics of dipole configurations even for moderate 
values of the chemical potential. We have also discussed the potential limitations
of the model in regimes of low chemical potential, or, in some cases,
small enough aspect ratios.
The balance between vortex-vortex interactions and the curvature effects
gives rise to four different types of dipole solutions: 
(i) the vertical dipole-in,
(ii) the horizontal dipole-in (vortices close to the inside part of the torus),
(iii) the horizontal dipole-out (vortices close to the outside part of the torus),
and (iv) the diagonal dipole.
The vertical-in and horizontal-out dipoles are (neutrally) stable for a wide range 
of parameters while the horizontal-out and diagonal dipoles are chiefly unstable.
The source of the relevant (stability or) instability via an eigendirection
associated with the relative motion of the vortices has been
identified and the related unstable dynamics also elucidated.
Nonetheless, for thick tori (large torus aspect ratio) it is possible to
stabilize the diagonal vortex dipole for sufficiently large chemical potentials.

We also explore vortex quadrupole configurations.
We found 16 different quadrupoles ranging
from horizontal and vertical aligned quadrupoles, 
to rectangular and rhomboidal quadrupoles,
to trapezoidal quadrupoles and, even some irregular quadrupoles.
All these solutions are continued and monitored for stability as the
torus aspect ratio is varied within the reduced ODE model.
Out of these 16 quadrupoles we found two
that exhibited windows of stability, upon variations of the aspect
ratio. We also found a handful of 
stationary quadrupole solutions with very weak instabilities indicating that 
it may be possible to observe them as long-lived solutions at the full NLS level.
Relevant stability considerations were also presented at the PDE
level and indeed, surprisingly it was found that some configurations,
such as the rectangular Q2 could be more robust at the latter level
and indeed genuinely stable for sizeable parametric intervals therein.
Finally, we briefly touch upon dark soliton stripe configurations.
These stripes can be single or double (or triple, etc.), horizontal or
vertical, and be centered about the inner or outer side of the torus.
Particularly interesting are the bifurcations of vortex configurations from 
these stripes as $\mu$ increases from the low density limit in a 
series of bifurcating cascades corresponding to increasing number of vortices,
reminiscent of ones emerging for stripe, as well as ring configurations
in regular two-dimensional parabolically confined BECs.

A natural extension of this work would be to study vortex configurations
with higher number of vortices. It would be indeed interesting to see if
configurations for higher number of vortices can be rendered stable for
the right parameter windows in a manner akin to what we detected for 
a couple of the quadrupole configurations (Q1 and Q3). It is likely
that configurations with higher number of vortices will be
difficult to be stabilized. Indeed, we encountered some such
configurations transiently in our dynamics (e.g., stemming from
unstable stripes).
It would also be interesting to understand in more detail how some full NLS
solutions seem to be more stable than their effective ODE model counterparts
(cf.~Q2 quadrupole).
On the other hand, it would also be relevant to study in more detail the 
bifurcation cascades between the different stripes ---single or double 
(or triple, etc.), in or out, vertical or horizontal--- and stationary 
vortex configurations. This is an intricate endeavor as these cascades
are very much dependent on the aspect ratio of the torus ($\alpha$).
For instance, a very thin torus ($\alpha\rightarrow 0$) will preferentially
promote the merger of vortices along the poloidal direction. However,
when the two torus radii are similar ($\alpha\rightarrow 1$) both
mergers along toroidal and poloidal directions will nontrivially compete.
Finally, further leveraging of the reduced ODE model could be employed to
study the existence and stability of periodic vortex orbits in a manner
akin to vortex choreographies in the plane~\cite{Calleja}. Indeed, an
example of such periodic solutions, returning to themselves upon
running around the torus are traveling solutions, such as the lump
ones spontaneously encountered herein. Given their potential connection
to so-called KP-lumps~\cite{nonlin}, this is an interesting direction
in its own right. Indeed, given the success of the particle model
herein, exploring additional directions such as the potential ordered
and chaotic orbits~\cite{skokos} at the low-dimensional dynamical systems
level could also hold some appeal. Lastly, we hope that this fruitful
comparison of ODE and PDE dynamics at the level of the torus will
springboard further related comparative studies in the context
of other nontrivial geometric settings, including spherical,
cylindrical and conical shells, among others.

\acknowledgements

J.D.A. gratefully acknowledges computing support based on the 
Army Research Office ARO-DURIP Grant W911NF-15-1-0403.
R.C.G.~gratefully acknowledges support from the US National Science Foundation
under Grants PHY-1603058 and PHY-2110038.
This material is based upon work supported by the US National 
Science Foundation under Grants DMS-1809074, PHY-2110030 (P.G.K.).



\begin{thebibliography}{60}

\bibitem{stringari}
L.P.~Pitaevskii and S.~Stringari,
{\it Bose-Einstein Condensation and Superfluidity},
Oxford University Press (Oxford, 2016).

\bibitem{pethick}
C.J. Pethick and H. Smith,
{\it Bose-Einstein condensation in dilute gases}, Cambridge University
Press (Cambridge, 2002).


\bibitem{siambook}
P.G.~Kevrekidis, D.J.~Frantzeskakis and R.~Carretero-Gonz\'alez,
{\it The Defocusing Nonlinear Schr{\"o}dinger Equation},
SIAM (Philadelphia, 2015). 

\bibitem{tsatsos}
See, e.g., N.P. Proukakis,
{\it Beyond Gross-Pitaevskii Mean-Field Theory}
in P.G. Kevrekidis, D.J. Frantzeskakis and R. Carretero-Gonz{\'a}lez,
{\it Emergent Nonlinear Phenomena in Bose-Einstein Condensates:
Theory and Experiment}, Springer-Verlag, Heidelberg (2008), p. 353.


\bibitem{experiment2}
L. Khaykovich, F. Schreck, G. Ferrari, T. Bourdel, J. Cubizolles, L.D. Carr, Y. Castin and C. Salomon,
Science \textbf{296}, 1290 (2002). 


\bibitem{expb1}
K.E.\ Strecker, G.B.\ Partridge, A.G.\ Truscott and R.G.\ Hulet, 
Nature \textbf{417}, 150 (2002).

\bibitem{expb2}
L.\ Khaykovich, F.\ Schreck, G.\ Ferrari, T.\ Bourdel, J.\ Cubizolles, L.D.\ Carr,
Y.\ Castin and C.\ Salomon,
Science \textbf{296}, 1290 (2002).

\bibitem{expb3}
S.L. Cornish, S.T. Thompson and C.E. Wieman,
Phys.\ Rev.\ Lett.\ {\bf 96}, 170401 (2006).


\bibitem{experiment}
S. Burger, K. Bongs, S. Dettmer, W. Ertmer, K. Sengstock, A. Sanpera, G.V. Shlyapnikov and M. Lewenstein,
Phys. Rev. Lett. \textbf{83}, 5198 (1999). 

\bibitem{experiment1}
J. Denschlag, J.E. Simsarian, D.L. Feder, Charles W. Clark, L.A. Collins, J. Cubizolles, L. Deng, E.W. Hangley, K. Helmerson, W.P. Reinhardt, S.L. Rolston, B.I. Schneider and W.D. Phillips,
Science \textbf{287}, 97 (2000). 


\bibitem{becker}
C.~Becker, S.~Stellmer, P.~Soltan-Panahi, S.~D{\"o}rscher, 
M.~Baumert, E.-M.~Richter, J.~Kronj\"{a}ger, K.~Bongs and K.~Sengstock, 
Nature Phys., {\bf 4}, 496
(2008).


\bibitem{markus}
A. Weller, J.P. Ronzheimer, C. Gross, J. Esteve,
M.K. Oberthaler, D.J. Frantzeskakis, G. Theocharis and P.G. Kevrekidis,
Phys.\ Rev.\ Lett.\ {\bf 101}, 130401 (2008).



\bibitem{engels}
P. Engels and C. Atherton, 
Phys. Rev. Lett. {\bf 99}, 160405 (2007).


\bibitem{becker2}
S. Stellmer, C. Becker, P. Soltan-Panahi, E.-M. Richter, S. D\"orscher, M. Baumert, J. Kronj{\"a}ger, K. Bongs and K. Sengstock,
Phys. Rev. Lett. {\bf 101}, 120406 (2008).

\bibitem{markus2}
G. Theocharis, A. Weller, J.P. Ronzheimer, C. Gross,
M.K. Oberthaler, P.G. Kevrekidis and D.J. Frantzeskakis, 
Phys.\ Rev.\ A {\bf 81}, 063604 (2010).

\bibitem{djf}
D.J. Frantzeskakis, J. Phys. A {\bf 43}, 213001 (2010). 

\bibitem{gap}
O. Morsch and M. Oberthaler,
Rev. Mod. Phys. \textbf{78}, 179 (2006).
  
  



\bibitem{revip}
P.G. Kevrekidis and D.J. Frantzeskakis, Rev. Phys. {\bf 1}, 140 (2016).


\bibitem{fetter1}
A.L. Fetter and A.A. Svidzinsky,
J. Phys.: Cond. Mat. {\bf 13}, R135 (2001).

\bibitem{fetter2}
A.L. Fetter, 
Rev. Mod. Phys. {\bf 81}, 647 (2009).

\bibitem{jeff}
I. Shomroni, E. Lahoud, S. Levy and J. Steinhauer,
Nat. Phys. {\bf 5}, 193 (2009).
  
\bibitem{komineas_rev}
S. Komineas,
Eur. Phys. J.- Spec. Topics {\bf 147}, 133 (2007).


\bibitem{lahnert}
K. Padavi{\'c}, K. Sun, C. Lannert, S. Vishveshwara,
Phys. Rev. A {\bf 102}, 043305 (2021). 

\bibitem{fetter3}
S. Bereta, M.A. Caracanhas, A.L. Fetter,
Phys. Rev. A {\bf 103}, 053306 (2021).

\bibitem{tononi}
A. Tononi and L. Salasnich,
Phys. Rev. Lett. {\bf 123}, 160403 (2019).

\bibitem{fetter:cylinder}
N.-E. Guenther, P. Massignan, and A.L. Fetter, 
Phys. Rev. A {\bf 96}, 063608 (2017).

\bibitem{fetter:cone}
P. Massignan and A.L. Fetter, 
Phys. Rev. A {\bf 99}, 063602 (2019).

\bibitem{pknewton}
P.K. Newton, {\it The N-Vortex Problem} (Springer-Verlag, 2001).

\bibitem{vitelli}
A.M. Turner, V. Vitelli, and D.R. Nelson,
Rev. Mod. Phys. 82, 1301 (2010).

\bibitem{lenn26}
N. Lundblad, R.A. Carollo, C. Lannert, M.J. Gold, X. Jiang,
D. Paseltiner, N. Sergay, and D.C. Aveline,
npj Microgravity {\bf 5}, 30 (2019).

\bibitem{lenn27}
S. Abend, W. Bartosch, A. Bawamia, D. Becker, H. Blume, 
C. Braxmaier, S.-W. Chiow, M.A. Efremov, W. Ertmer,
P. Fierlinger, N. Gaaloul, J. Grosse, C. Grzeschik, O. Hellmig,
V.A. Henderson, W. Herr, U. Israelsson, J. Kohel, M. Krutzik,
C. K\"urbis, C. L{\"a}mmerzahl, M. List, D. L{\"u}dtke, N. Lundblad,
J.P. Marburger, M. Meister, M. Mihm, H. M{\"u}ller, H. M{\"u}ntinga,
T. Oberschulte, A. Papakonstantinou, J. Perovsek, A. Peters,
A. Prat, E.M. Rasel, A. Roura, W.P. Schleich, C. Schubert,
S.T. Seidel, J. Sommer, C. Spindeldreier, D. Stamper-Kurn,
B.K. Stuhl, M. Warner, T. Wendrich, A. Wenzlawski, A. Wicht,
P. Windpassinger, N. Yu, and L. W{\"o}rner,
EPJ Quantum Technology {\bf 8}, 1 (2021).


\bibitem{garraway}
B.M. Garraway and H. Perrin, J. Phys. B: At. Mol. Opt. Phys. {\bf 49}, 172001 (2016).



\bibitem{porto}
H. Kim, G. Zhu, J.V. Porto and M. Hafezi
Phys. Rev. Lett. {\bf 121}, 133002 (2018).


\bibitem{fetter:torus}
N.-E. Guenther, P. Massignan, A.L. Fetter,
Phys. Rev. A {\bf 101}, 053606 (2020).



\bibitem{torus:bright}
J. D'Ambroise, P.G. Kevrekidis, P. Schmelcher,
Phy. Lett. A {\bf 384}, 126167 (2020).


\bibitem{glowinski}
R. Glowinski and D.C. Sorensen,
in {\it Partial Differential Equations: Modeling and Numerical Simulation}, 
R. Glowinski and P. Neittaanm{\"a}ki (Eds),
Springer-Verlag (Berlin, 2008) p. 225.



\bibitem{stremler:JFM99}
M.A. Stremler and H. Aref,
J. Fluid Mech. {\bf 392}, 101
(1999).


\bibitem{kirchhoff}
G. Kirchhoff, \"Uber die station\"aren elektrischen Str\"omungen in
einer gekr\"ummten leitenden Fl\"ache, 
Monatsber. Akad. Wiss. Berlin, 487 (1875).

\bibitem{Middelkamp:2011}
S. Middelkamp, P.G. Kevrekidis, D.J. Frantzeskakis, R. Carretero-Gonz\'alez, and P. Schmelcher.
Physica D {\bf 240}, 1449 (2011).
    
\bibitem{Jones-Robert:1982}
C. Jones and P. Roberts,
Motion in a Bose condensate IV. Axisymmetric solitary waves.
J.\ Phys.\ A: Math.\ Gen.\ {\bf 15}, 2599 (1982).


\bibitem{bongs}
N. Meyer, H. Proud, M. Perea-Ortiz, C. O'Neale, M. Baumert, M. Holynski, J. Kronj{\"a}ger, G. Barontini, and K. Bongs,
Observation of Two-Dimensional Localized Jones-Roberts Solitons in Bose-Einstein Condensates,
Phys.\ Rev.\ Lett.\ {\bf 119}, 150403 (2017).


\bibitem{Calleja}
R.C. Calleja, E.J. Doedel, and C. Garc\'{\i}a-Azpeitia,
Choreographies in the $n$-vortex Problem. 
Regul.\ Chaot.\ Dyn.\ {\bf 23}, 595 (2018).


\bibitem{nonlin}
D. Chiron and C. Scheid, 
Nonlinearity {\bf 31}, 2809 (2018).

\bibitem{skokos}
See, e.g., 
N. Kyriakopoulos, V. Koukouloyannis, C. Skokos, P.G. Kevrekidis,
Chaos {\bf 24}, 024410 (2014).

\end{thebibliography}
\end{document}